\documentclass[twocolumn]{aastex631}

\usepackage{float}
\usepackage{hyphenat}
\usepackage{graphicx}
\usepackage{xcolor}
\usepackage{color}
\usepackage[flushleft]{threeparttable}
\usepackage{comment}

\newcommand{\argmax}{\textrm{argmax}}
\newcommand{\argmin}{\textrm{argmin}}  

\newcommand\DEFAULTFIGWIDTH{.48} 
\newcommand\thirdimgwidth{.45}

\begin{document}

\shorttitle{Classifying Transients Using Host Galaxy Photometry}
\shortauthors{Kisley et al.}
\title{Classifying Astronomical Transients Using Only Host Galaxy Photometry}
\author[0000-0003-4573-8095]{Marina Kisley} 
\affiliation{Department of Computer Science, University of Arizona}
\author[0000-0003-3658-6026]{Yu-Jing Qin} 
\affiliation{Department of Astronomy, University of Arizona}
\author[0000-0001-6047-8469]{Ann Zabludoff}
\affiliation{Department of Astronomy, University of Arizona}
\author[0000-0002-8568-9518]{Kobus Barnard}
\affiliation{Department of Computer Science, University of Arizona}
\author[0000-0003-2158-8141]{Chia-Lin Ko}
\affiliation{Department of Astronomy, University of Arizona}

\begin{abstract}
The Legacy Survey of Space and Time (LSST) at the Vera C. Rubin Observatory will discover tens of thousands of extragalactic transients each night. The high volume of alerts demands immediate classification of transient types in order to prioritize observational follow-ups before events fade away. We use host galaxy features to classify transients, thereby providing classification upon discovery. In contrast to past work that focused on distinguishing Type Ia and core-collapse supernovae (SNe) using host galaxy features that are not always accessible (e.g., morphology), we determine the relative likelihood across $12$ transient classes based on only 19 host apparent magnitudes and colors from $10$ optical and IR photometric bands.
We develop both binary and multiclass classifiers, using kernel density estimation to estimate the underlying distribution of host galaxy properties for each transient class. Even in this pilot study, and ignoring relative differences in transient class frequencies, we distinguish eight transient classes at purities significantly above the 8.3\% baseline (based on a classifier that assigns labels uniformly and at random):
tidal disruption events ($48\%\pm27\%$, where $\pm$ indicates the 95\% confidence limit),
SNe Ia-91bg ($32\%\pm18\%$),
SNe Ia-91T ($23\%\pm11\%$),
SNe Ib ($23\%\pm13\%$),
SNe II ($17\%\pm2\%$),
SNe IIn ($17\%\pm6\%$), 
SNe II P ($16\%\pm4\%$), and
SNe Ia ($10\%\pm1\%$). 
We demonstrate that our model is applicable to LSST and estimate that our approach may accurately classify 59\% of LSST alerts expected each year for SNe Ia, Ia-91bg, II, Ibc, SLSN-I, and tidal disruption events. 
Our code
\footnote{\url{https://github.com/marinakiseleva/thex_model}, \url{https://github.com/marinakiseleva/z_dist}} and
dataset\footnote{\url{https://sandbox.zenodo.org/record/1086145}.} are publically available.
\end{abstract} 
\section{Introduction} \label{sec:intro}

Transient astronomical events, referred to as transients, are intense, bright, and short-lived phenomena that briefly light up the sky before fading away. 
Most extragalactic transients are related to stellar deaths and their remnants. These include core-collapse and Type Ia supernovae, as well as their subtypes, gamma-ray bursts, kilonovae, and tidal disruption events (TDEs), where a star falling into a galaxy's central black hole is torn apart \citep[e.g.,][]{Gezari2021}.

The extreme conditions of these events provide means to study theories of general relativity and cosmology \citep[e.g.,][]{Goobar2011, 2013arXiv1305.5720E, Demianski2017, Mockler2019}.
Discovering and classifying transients are priorities for the Vera C. Rubin Observatory, which in 2024 is expected to capture tens of thousands of extragalactic
transients each night during its Legacy Survey of Space and Time \citep[LSST;][]{2009arXiv0912.0201L, Ivezic2019}. LSST will generate  alerts to enable some transients to be followed\hyp{}up with other telescopes.

LSST will detect transients at an unprecedented rate, exceeding our capabilities to follow-up most of them, especially with spectroscopic and/or multi-wavelength observations. Currently, only one-tenth of optical transients can be classified with spectroscopic follow-up \citep{Kulkarni2020}, and the gap will widen by orders of magnitude in the LSST era. Selecting the most interesting candidates for each science case quickly, to capture the early-stage evolution and before the transient fades away, is essential.

We focus here on the classification of transient events using the properties of their host galaxies, which are often known beforehand. 
The connections of transient types to specific stellar progenitor systems imply that the relative frequencies of transients depend on the properties of the stellar populations in their host galaxies \citep[e.g., ][]{Oemler1979, Cappellaro1988, Li2011, Graur2017}. Such dependencies should make it possible to classify transients at detection, well before their spectra or full light curves become available.
In this way, our approach preemptively classifies transients by considering the most likely transient to occur given a particular galaxy's properties. 

Most approaches to transient classification do not use host galaxy properties as input features, but instead classify based on the transient light curve. Such methodologies are effective, yet require days, weeks, or even longer to collect data \citep{2018arXiv181000001T, Kessler_2019, Muthukrishna_2019, boone2019avocado, Neira2020, Villar2020, Burhanudin2021, Qu2021, Qu2022}. 
In the case of LSST, the average time between revisiting the same object is three days \citep{Ivezic2019}. Given the delay in classification when using light curves, we focus our attention on  classification using host galaxy data, which allows for instantaneous follow-up upon discovery, permitting us to glimpse rare or quick events before they fade away.  

Although previous work has empirically demonstrated that certain transient classes favor certain galaxies, and may be distinguished at significant rates using host galaxy features, few address the problem in context of LSST or with more than two classes \citep{2016AAS...22831402A, pan2014host, French_2018}. 
Previous research in distinguishing Ia and core-collapse host galaxies often relied on certain physical properties of galaxies, such as morphology and luminosity, which are not going to be as widely available as apparent magnitude for galaxies observed by LSST \citep{Foley_2013, Gagliano:2020ucg}.

We attempt to fill this gap in transient classification by developing an LSST-applicable model to distinguish among $12$ transient classes, using only $10$ host galaxy photometric magnitudes (and nine derived colors).  
We pose the question of classifying transient types as two distinct scientific cases. We first address the question of determining the probability that the observed event is of a certain class. This methodology constitutes our binary classifier approach in Section \ref{sec:binarymodel}. Although this provides the independent probability of each transient class, the resulting probabilities across the range of classes are not comparable. To address the question of the most likely transient type for an event, we develop multiclass classifiers, which provide the probabilities across the range of transients under consideration, and classify them by the maximum assigned probability (\S \ref{sec:ovaclassifier} and \ref{sec:multimodel}).  

In Sections \ref{sec:binary_results} and \ref{sec:multi_results}, we evaluate our methods in terms of classification performance. We measure the classification performance by considering how accurate the model is in predictions (purity) as well as the proportion of events we can expect to accurately capture (completeness). In Section \ref{sec:prob_performance}, we demonstrate that the purity of our methods may be significantly improved when using the probabilities assigned to events. In Section \ref{sec:light_curves}, we compare our model to alternative approaches that use transient light curves or other host galaxy features. In Section \ref{sec:Zsection}, we evaluate our method's applicability to LSST. 
To ensure that our method provides meaningful likelihoods across a range of transient classes for a significant portion of LSST alerts, we employ data distributed similarly to the anticipated transient detections of LSST. In particular, we address the systematic biases in our data versus those of the anticipated LSST data with respect to redshift.

\begin{figure}
    \centering
    \includegraphics[width=0.5\textwidth]{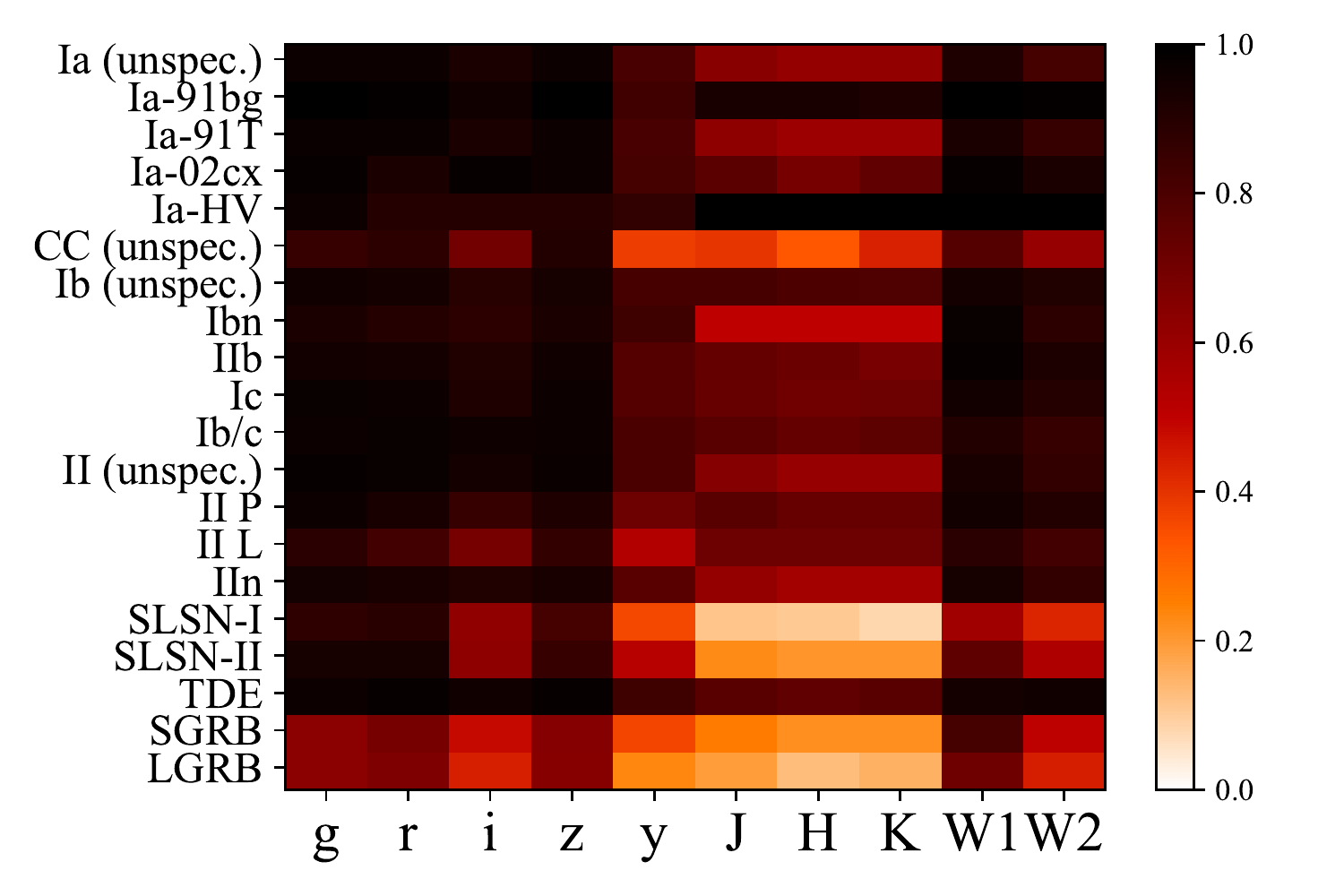}
    \caption{
    Fraction (colorbar) of host galaxies from \citet{Qin2021}
    in each transient class (vertical axis) with valid (SNR $> 3$, unflagged) apparent magnitudes in each of the 10 photometric bands along the horizontal axis.
    The other nine features that we use as inputs to our transient classification  analysis are colors derived from these 10 photometric bands.
    The inconsistency of feature availability across different classes is due to the fact that the
    \citet{Qin2021} database is a combination of different catalogs that have been merged together.  
    Details of the transient class hierarchy are presented in \S \ref{sec:classtreatment}.
    }
    \label{fig:allfeaturescompleteness}
\end{figure}

\section{Data} \label{sec:data}

We ensure the proposed methodology is pertinent to the LSST by using data that already exists or that we can expect to obtain for most galaxies observed by LSST. Future LSST data can be incorporated into our proposed model to classify transients for potential host galaxies that previously lacked photometry. We focus here on photometric data ranging from visible to infrared (IR). Many galaxies observed by LSST (on the order of millions) will have optical and IR data available from previous surveys, and LSST itself will collect light for six optical passbands: $u, g, r, i, z, y$.  

Below we describe how we construct the ``THEx'' (Transient Host Exchange) dataset\footnote{ doi:10.5072/zenodo.1086145.} used here for training and testing (\S \ref{sec:datasetcreation}) 
and how we treat the hierarchy of supernova subtypes (\S \ref{sec:classtreatment}).
The dataset is drawn from the \citet{Qin2021} database and consists of well-matched host-transient pairs that have a complete set of high-quality host optical-IR photometric magnitudes. For hosts with multiple magnitude measurements in the same photometric band from different surveys, we assign single magnitudes after cross-calibrating across surveys to achieve a common magnitude system.
Section \ref{sec:mergingcatalogs} describes how photometric magnitudes are cross-calibrated across surveys and assigned to host galaxies.
Section \ref{sec:hosttransientmatching} details how the best-matched host-transient pairs are selected.

\begin{figure}
    \centering
    \includegraphics[width=\DEFAULTFIGWIDTH\textwidth]{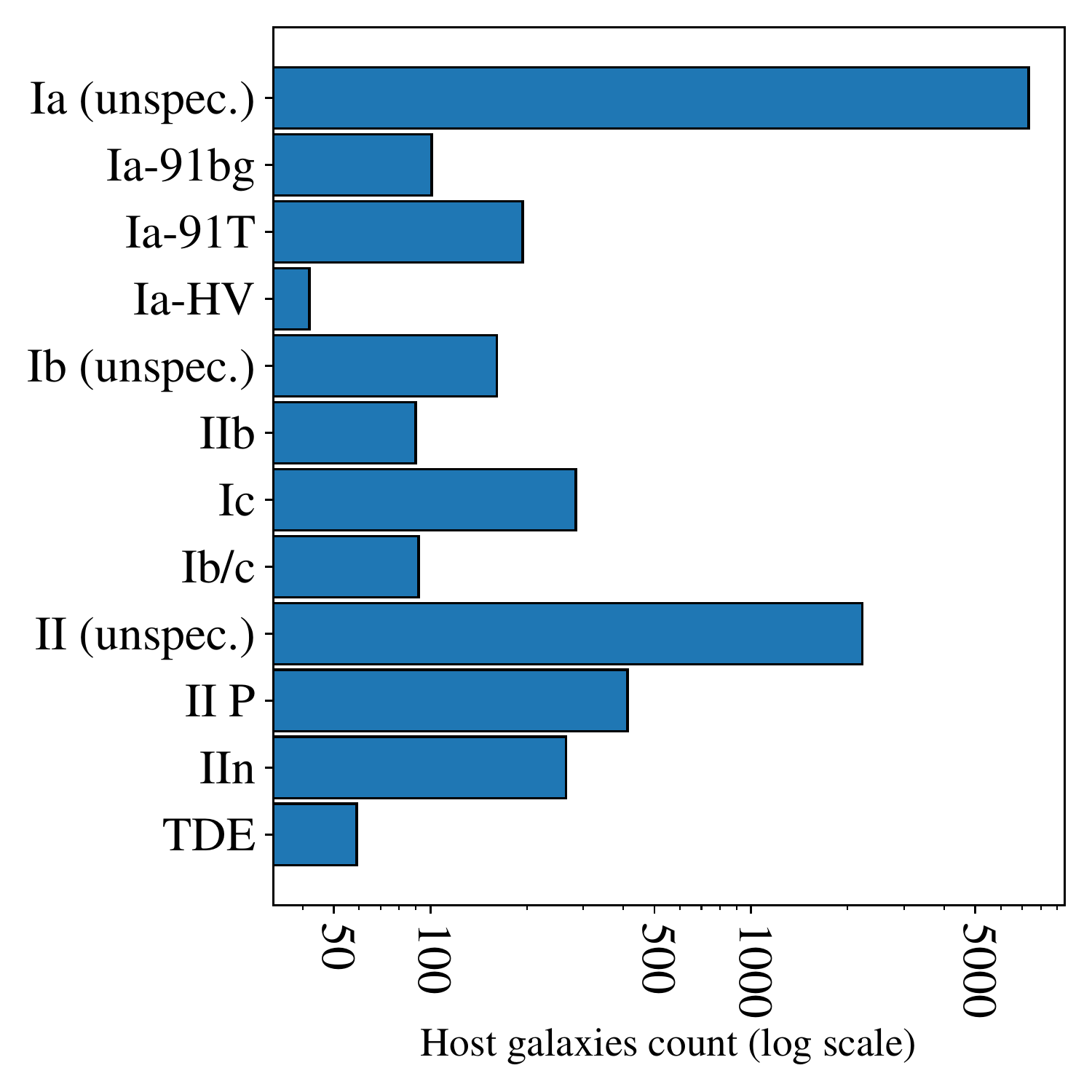}
    \caption{Host-transient pair sample sizes for the 12 transient classes with at least 40 hosts with all 19 photometric features: 10 magnitudes ($g, r, i, z, y, J, H, K, W1, W2$) plus the nine associated colors. Classes removed due to inadequate data include Ia-02cx, CC (unspec.), Ibn, II L, SLSN-I/II, and SGRB/LGRB. The largest classes are Ia (unspec.) and II (unspec.). 
    In total, we have 11260 host\hyp{}transient pairs in our dataset. 
   }
    \label{fig:classdist_g_W2}
\end{figure}

\subsection{THEx Dataset}
\label{sec:datasetcreation}

We use broadband photometric magnitudes (and the associated colors) as host galaxy features for our  transient classifiers.
Broadband  magnitudes are tracers of the spectral energy distributions (SEDs) of galaxies and are more widely available than spectroscopic or morphological data.
The database of \citet{Qin2021} is a comprehensive compilation of host galaxy optical and IR  magnitudes and has over $38,000$ unique host-transient pairs.
The magnitudes are collected from numerous galaxy surveys\footnote{
The Sloan Digital Sky Survey (SDSS) DR16 \citep{Ahumada2020}, DESI Legacy Imaging Surveys DR8 \citep{Dey2019},  Panoramic Survey Telescope and Rapid Response System (Pan-STARRS) DR2 \citep{Chambers2016}, Dark Energy Survey DR2 \citep{Abbott2018, Abbott2021},  Two-micron All-Sky Survey (2MASS) Point and Extended Source Catalogs \citep{Jarrett2000, Skrutskie2006},  Large Area Survey of the UKIRT Infrared Deep Sky Survey (UKIDSS-LAS) DR9 \citep{Lawrence2007}, VISTA Hemisphere Survey DR4 \citep{McMahon2013}, and  SkyMapper Southern Survey DR2 \citep{Keller2007}.}.

The supernova types and tidal disruption events in \citet{Qin2021} are mostly determined spectroscopically. Only the SDSS\hyp{}II dataset \citep{Campbell13} in our original database employs photometric classification of SNe Ia. That particular catalog contributes only 352 SNe to the original database ($<1$\% of all events); most of these are at higher redshifts where the host galaxy photometry is incomplete and therefore excluded from our subsequent analyses.
Thus, the transient classes we use are generally spectroscopic and should be accurate. Below we discuss the supernova subtypes in more detail.

Progenitors of core-collapse supernovae (CC SNe) are generally thought to be massive stars \citep{Smartt2009}. Depending on the degree of pre-explosion mass loss, these events may give rise to vastly different phenomena, including hydrogen-rich Type-II supernovae, which usually feature a plateau in the post-maximum light curve (Type II-P) \citep{Pian2017}, as well as hydrogen-deficient or stripped-envelope (SE), Type Ib, Ic, or the transitional Type IIb supernovae \citep{Arcavi2017}. Driven by various mechanisms such as pair-instability supernovae, shock interactions with circumstellar material (CSM), or magnetars, CC SNe with extremely high luminosities are also classified as superluminous supernovae \citep[SLSN;][]{Gal-Yam2019}.

Type Ia supernovae, on the other hand, arise from the thermonuclear explosions of white dwarfs \citep{Maoz2014}. Despite being homogeneous in observed properties, there are several minor subgroups, including SN Ia-1991T-like and SN Ia-1991bg-like (hereafter Ia-91T and Ia-91bg), which may represent diverse progenitor properties and channels \citep{Taubenberger2017}. Based on the maximal-light spectra, there is also a subgroup with high velocity silicon lines \citep[Ia-HV;][]{Branch2006, Wang2009, Blondin2012}. For both core-collapse and thermonuclear supernovae, shock interaction with the CSM may lead to peculiar subtypes like Type Ibn and IIn \citep{Smith2017}.

The THEx dataset consists of optical and IR photometric magnitudes of host galaxies and their respective colors.
The 19 photometric features, including 10 photometric bands and nine derived colors, used later in our models are: $g, g-r, r, r-i, i, i-z, z, z-y, y, y-J, J, J-H, H, H-K, K, K-W1, W1, W1-W2,$ and $W2$, where $W1$ and $W2$ are mid-IR passbands from the \textit{Wide-field Infrared Survey Explorer} \citep[\textit{WISE}; ][]{Cutri2014}. Our dataset is limited only to valid magnitudes, i.e., those with a signal-to-noise ratio (SNR) of $>$ 3 and without any flags indicating poor quality in their original surveys, as outlined in Section 4 of \citet{Qin2022}.

Figure \ref{fig:allfeaturescompleteness} visualizes our dataset in terms of its completeness across the 10 photometric bands.
Figure \ref{fig:classdist_g_W2} displays the host galaxy sample sizes in each of the 12 transient classes that remain after we remove those classes with inadequate data for training.
We conduct our analyses on only those classes with at least $40$ host galaxies (with the exception of \S \ref{sec:Zsection}, which incorporates smaller classes).
We remove the classes Ia-02cx, CC (unspec.), Ibn, II L, SLSN-I/II, and LGRB/SGRB, because they generally lack $y, J, H, $ and $K$ host magnitudes.

\subsubsection{Photometric Magnitudes}
\label{sec:mergingcatalogs}

The \citet{Qin2021} database does not include all 10 photometric magnitudes for each transient host (see Figure \ref{fig:allfeaturescompleteness}). Furthermore, for some hosts, there are multiple measurements of a magnitude in the same band from different surveys. To address these problems, we first downselect our sample to include only those hosts with all 10 photometric magnitudes. Then, we cross-calibrate where there are multiple measurements in the same band for a given host to effectively place all magnitudes on the same system.

Different photometric systems, such as the AB system, Vega system, or SDSS arcsinh system, may have different definitions of magnitude, zero points for bands, and calibration standards \citep{1999AJLupton}.  Surveys also differ in their filter profiles and instrumental response. Differences in photometric techniques (aperture photometry versus profile-fitting photometry, and the associated parameters) can lead to systematically different measurements. When combining multiple magnitudes into one, instead of choosing a “best” magnitude or merely taking the average or median value, we choose one photometric catalog in each band as the standard and, when necessary, calibrate other measured magnitudes to this standard to reduce systematic offsets between different surveys.

For each band, we identify a target magnitude column in the original database, selected based on which has the best data availability for our host galaxies. This magnitude column is the preferred one when multiple magnitudes are reported in this band. If the target magnitude of a band is available, we use that value as-is. Otherwise, we use another available magnitude value in the same nominal band (source magnitude), with proper cross-calibration to minimize the systematic offsets between the selected source magnitude column and our target magnitude column. For example, we use the $g$-band Kron magnitude from the Pan-STARRS catalog, which is the target magnitude column for this band, when it is available. If this target magnitude is not reported, but $g$-band magnitudes from other surveys are available, we choose one of those magnitudes as the source magnitude and calibrate it to our target magnitude.

To cross-calibrate the source and target magnitude, we use linear regression to minimize the differences between the magnitudes and a secondary set of properties.
We assume that the systematic offset of source magnitude ($m_s$) and target magnitude ($m_t$) depends on a secondary property ($x$) in a linear form: $m_t = m_s + k x + c$, where $x$ is a column in the source magnitude catalog, $k$ is the linear regression slope, and $c$ is a constant offset. To calibrate $m_s$ to $m_t$, we first use hosts with measured $m_s$ and $m_t$ to fit $k$ and $c$, under all possible choices of $x$. We use Orthogonal Distance Regression \citep[][implemented in \texttt{scipy}]{Boggs1989}, which allows us to take the error of $m_t$ and $m_s$ into account and to fit $k$ and $c$ coefficients here. 
We then choose column $x$ with the minimal median error in $kx + c$. This secondary property $x$, in combination with the associated coefficients $k$ and $c$, calibrates $m_s$ to $m_t$. Usually, the magnitude in another band, color, or a shape parameter is chosen as the secondary property here.
Overall, this process aims to reduce the uncertainties in the magnitude values and optimizes the transformation between the two magnitude columns. 

In the event there are multiple possible source magnitudes, each one is transformed into the target magnitude using the pre-calculated cross-calibration of each survey. We then use the source magnitude that has the least mapping error in the calibrated magnitude, considering both the error of original magnitude and the uncertainties in the linear regression.

\subsubsection{Selection of Host-Transient Pairs}
\label{sec:hosttransientmatching}

We use the visual inspection quality flags in \citet{Qin2021} to select a subset of reliable transient-host pairs.
For transients with previously known host galaxies, we select those for which the host name or coordinate cross-matches correctly with other catalogs (Case A1 in 
\citet{Qin2021}; $62\%$ of our training set) or for which the host is manually re-assigned to correct a likely mistake of cross-matching (Case B1; $<1\%$ of our training set).
For transients with newly identified hosts, we select those for which the algorithm-identified host appears reliable and properly cross-matched with other catalogs (Case F1, $37\%$ of our training set) or those for which the host is manually re-assigned due to a likely mistake in host association or cross-matching (Case G1, $1\%$ of our training set).

\begin{figure}
    \centering
    \includegraphics[width=\DEFAULTFIGWIDTH\textwidth]{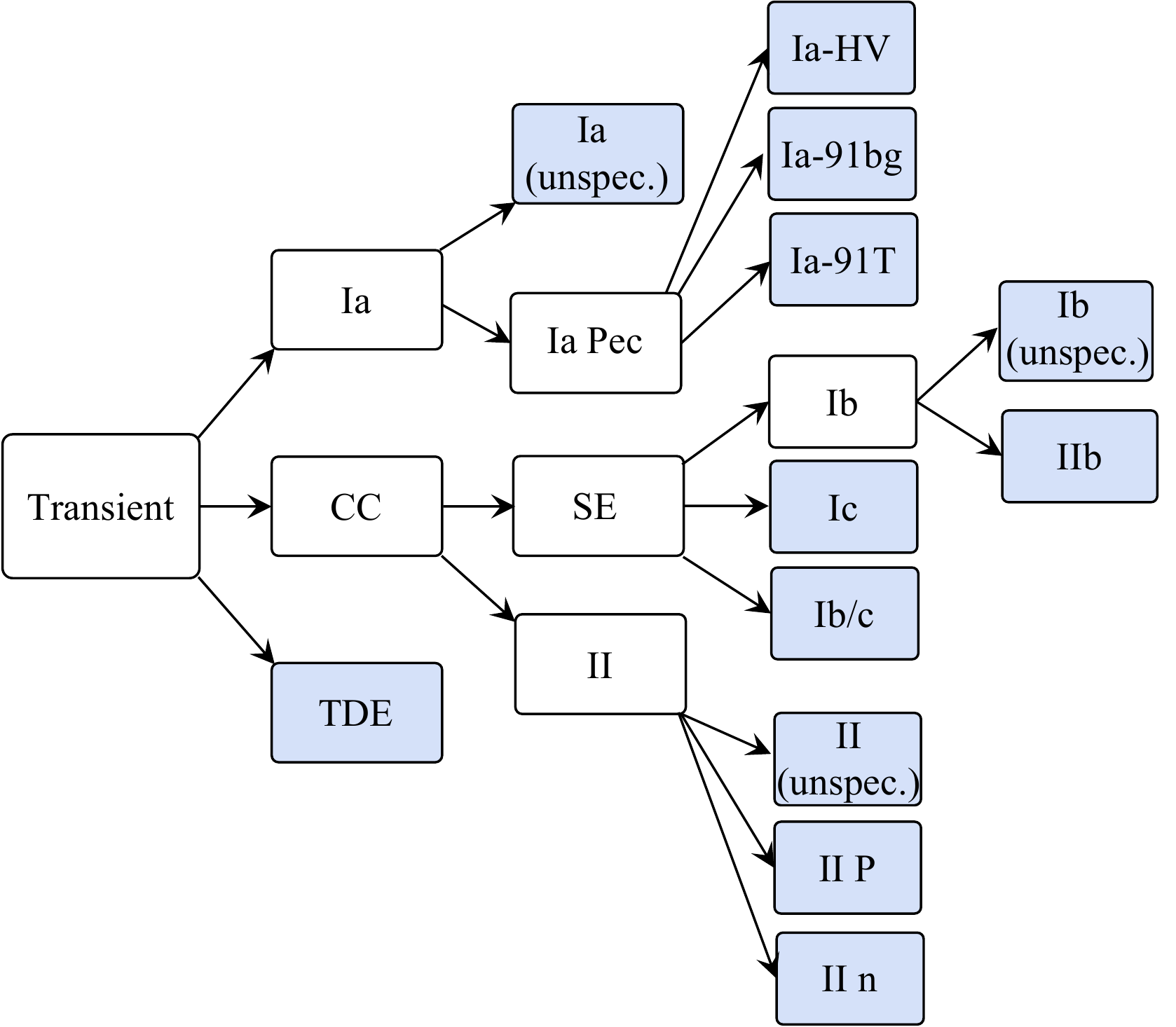}
    \caption{The transient class hierarchy, where \textit{Transient} is the root. The classes shaded in blue correspond to the lowest\hyp{}level, unique transient classification available. Each host\hyp{}transient pair in the dataset may have multiple class labels across different levels of the hierarchy (e.g., Ia, Ia Pec, and Ia-91bg). Additionally, samples may have labels that terminate at a higher level of the hierarchy. \textit{Unspecified} class labels, represented as \textit{unspec.}, correspond to samples whose lowest\hyp{}level classification is the parent class. The class labels shaded in blue form a disjoint set over which we may compute the comparative probabilities of these events. }
    \label{fig:classhierarchy2}
\end{figure}

\subsection{Class Hierarchy Treatment} 
\label{sec:classtreatment}

Transient classes are related to one another in a hierarchy, depicted in Figure \ref{fig:classhierarchy2} (equivalent to that in \citet{Qin2021}). Transient events may have a label at a single level or across several levels of the class hierarchy. For example, a Ia-91bg event has the labels Ia, Ia Pec, and Ia-91bg. When determining the probability of each event, we ensure that probabilities are normalized over the disjoint set of events. To do so, we redefine labels such that each event is defined by its most detailed, lowest\hyp{}level class assignment in the hierarchy (denoted in blue in Figure \ref{fig:classhierarchy2}). In the case of Ia-91bg, we would consider the event only as Ia-91bg and not also as Ia. Alternatively, if an event's lowest classification is Ia, we will classify it as Ia (unspec.). \textit{Unspecified} is a term we use to label those host-transient pairs which do not have any lower\hyp{}level label associated with them. In the case of Ia,  Ia (unspec.) are those Type Ia events not associated with any Ia subtype considered here (Ia-91bg and Ia-91T). 

Obtaining a clean sample of Branch-normal SN Ia (i.e., not contaminated by subtypes) from archival surveys is challenging. As a result, our training set may not reflect the true relative frequencies of 91T, 91bg, and normal SN Ia as a function of host galaxy properties. For normal SN Ia, the impact should be limited, as both subtypes consist of a small fraction of all SN Ia. As revealed by the Berkeley Supernova Ia Program \citep{Silverman12}, about 6\% of all SNe Ia are 91bg and 2\% are 91T/99aa. Moreover, although these subtypes show spectroscopic and photometric properties distinct from the majority of normal SN Ia, Ia-91T/Ia-91bg represent  two ends of the continuum of SN Ia photometric properties. For these two subtypes, we achieve reasonable performance using our training data, which should be assessed with a more coherently classified transient sample in the future.

\section{Methodology}
\label{sec:methodology}
We aim to develop a model that is able to accurately identify the most likely transient to occur in a given host galaxy.
We compare three different approaches, one of which addresses the question in terms of binary classification, and two of which address the problem in terms of multiclass classification. In the binary case, the probability of each class is estimated using a unique classifier. This provides the likelihood of each class separately, which addresses the needs of researchers interested in only the independent probability of any single class, not in the most likely class overall. 
To address which transient class of the considered classes is the most likely, we develop two multiclass classifiers, which provide the relative probabilities across the range of classes. Due to the nature of transient classes and their hierarchical relationship to one another, we determine the multiclass probability based on a disjoint set of transient classes discussed in Section \ref{sec:classtreatment}. This disjoint set of events corresponds to mutually exclusive possibilities for a transient type. For example, we may consider the probability of Ia and TDE, but we may not compare the likelihood between CC, SE, and Ib, since an event may be all three.  

Because the anticipated relative frequencies of transient events detected by LSST are not well known for all classes considered, we focus here on the likelihood of events based on the data. This considers how similar a given galaxy appears to be to those of each class in the training data, but defers considering the frequency of the classes. For example, the fact that SNe Ia are more common than tidal disruption events is not considered in building the likelihood model. As such, our model may be updated to incorporate class frequencies as they become available. Section \ref{sec:incpriors} briefly explores the anticipated performance changes when incorporating prior probabilities based on class frequencies for a handful of classes whose rates are generally known for the LSST.

\subsection{Binary Classifiers} 
\label{sec:binarymodel}
We develop a unique binary classifier for each transient class, with the aim of differentiating that class from the rest of the dataset based on features of the host galaxy. The objective of our model is to provide information to the community on whether or not follow-up of a particular event is warranted. For this reason, we consider a probabilistic approach that provides the probability of an event belonging to each transient class. Using Bayes’ rule, we factor the probability of the event into its likelihood (based on the data) from the prior probability of the event (based on how frequently we expect to observe each class). Non-probabilistic but common approaches in this area, such as random forest classifiers or neural networks, do not make this distinction, and therefore are unable to distinguish uncertainty in the likelihood versus uncertainty in our prior knowledge of the class's frequency.  
 
We use $t_k$ to indicate whether a transient belongs to class $k$ ($t_k = 1$) or not ($t_k=0$). We find the probability of each transient class, $t_k$, using Bayes' theorem:
\begin{equation}
P_B(t_k=1|\textbf{x}) = \frac{p(\textbf{x}|t_k=1)}{p(\textbf{x}) } p(t_k=1) ,
\label{eq:binarybayesfull}
\end{equation}
where $\textbf{x}$ is a vector of galaxy features.
$p(t_k=1|\textbf{x})$ is the posterior probability of the transient type given the galaxy features, $p(\textbf{x}) = p(t_k=1)p(\textbf{x}|t_k=1)+p(t_k=0)P(\textbf{x}|t_k=0)$ is the evidence over $N=2$ classes (the class $t_k$ and the inverse, not class $t_k$), $p(t_k=1)$ is the prior probability of class $t_k$, and $p(t_k=0)$ is the prior probability of the sample not being class $t_k$. 
 
We focus on a likelihood-only model, which estimates the probability of an event's class based solely on how a galaxy's data resembles that of a particular class, (the likelihood, $p(\textbf{x}|t_k=1)$), generalized from the training data. We forego any assumptions on prior probability because the rates of identification by LSST for most of the transient classes considered here are not yet known. In Section \ref{sec:incpriors} we briefly explore the potential performance improvements from incorporating priors for five of the $12$ classes which have available rates. For the rest of this paper, we incorporate all $12$ classes into the model by focusing on only the likelihood of events. To create a likelihood-only model, we functionally ignore the prior probability, which is equivalent to assuming uniform prior probabilities across classes. As such, Equation \ref{eq:binarybayesfull} simplifies to:

\begin{equation}
P_B(t_k=1|\textbf{x}) = \frac{p(\textbf{x}|t_k=1)}{  p(\textbf{x}|t_k=1) + p(\textbf{x}|t_k=0) } .
\label{eq:binarybayes}
\end{equation}
  
The likelihood (the numerator above) is described by a multi-dimensional distribution in the parameter space: a unique distribution for each class, providing the likelihood for a set of features in that class, $p(\textbf{x}|t_k=1)$. Similarly, $p(\textbf{x}|t_k=0)$ is the likelihood of the event \textit{not} belonging to the class. Because the shape of these distributions is unknown, we use a common non-parametric density estimation technique known as  kernel density estimation (KDE), which allows us to estimate the unknown probability density distributions \citep{parzen1962estimation}. In particular, for each class, we use multivariate KDE, which estimates a distribution over all dimensions simultaneously. The density estimate at each point $x$ is given with respect to the local neighborhood of training points $x_i, i=1, ..., n$:

\begin{equation}
f(x)= \frac{1}{nh}\sum_{i=1}^n K(\frac{x - x_i}{h}),
\end{equation} 
where $h$ is the smoothing parameter known as the bandwidth, $n$ is the number of samples in the local neighborhood, and $K$ is the kernel function. Any smooth unimodal function with a peak at $0$ may be used as the kernel function, the most common example being the Gaussian kernel ($K(x;h) = \exp{\frac{-x^2}{2h^2}}$). There are no constraints on the local neighborhood of a point for the Gaussian kernel, so we use $n=N$, the total number of training points. 

We find the best-fitting kernel and bandwidth per class using a grid search. For each kernel, and each bandwidth in an acceptable range, we fit the estimator to a portion of the training data and evaluate on the remaining training data (the validation set).  
To determine how well the bandwidth and kernel describe the data, we estimate the misclassification loss on the validation set to promote distributional estimates that are good for classification, but potentially less good for characterizing the distribution. 
For misclassification error, we use the Brier score loss, 
\begin{equation}
    B(\textbf{x}, \textbf{y}; t_k) =\frac{1}{N} \sum_{i=1}^{N}\left(P(t_k = 1 | \textbf{x}_i ) - y_i \right)^{2},
    \label{eq:brierscore}
\end{equation}
where $P(t_k = 1 | \textbf{x}_i)$ is the probability of class $t_k$ for data point $x_i$, $y_i$ is the actual class of the data point ($y_i=1$ when the class is $t_k$ and $y_i=0$ otherwise), and $N$ is the number of data points being evaluated. The Brier score is equivalent to the mean squared error of the predictions and allows us to evaluate the accuracy of the probabilities. As with mean squared error, the smaller the Brier score, the better the model performs. 

The optimal bandwidth and kernel are identified as those with the smallest loss for the validation set:
\begin{equation}
    \hat{h}, \hat{l} = \argmin_{h \in H, l \in L}  B(\textbf{x}, \textbf{y}),
\end{equation}
where $\hat{h}, \hat{l}$ are the optimal bandwidth and kernel, respectively, $H$ is the entire range of bandwidths considered, $L$ is the entire range of kernels, and  $\textbf{x}, \textbf{y}$ is the validation dataset (30\% of the training data). For consistency, we use the same training/validation split for all kernels/bandwidths evaluated.  The range of bandwidths is dictated by the range of values in the dataset. The range of kernels considered include the standard kernel types: Gaussian, exponential, Epanechnikov, tophat, linear, and cosine. We find that the best fitting kernel across all classes is generally the exponential kernel, ($\hat{l}(x;h) = \exp{\frac{-x}{h}}$), followed by the Gaussian. 
Thus, we can compute the class probabilities (Equation \ref{eq:binarybayes}) through the likelihood function (using the exponential kernel for an example):
\begin{equation}
p(\textbf{x}|t_k=1) \propto \frac{1}{nh} \sum_{i=1}^n \exp(-\frac{||\textbf{x} - \textbf{x}_i||}{h}).
\end{equation}

In our model, we consider the Gaussian and exponential kernels for the binary and OVA classification and use the exponential kernel for the multiclass KDE classification. 
The best-fit kernels and the bandwidths used in our analysis are listed in Appendix \ref{sec:appendix_kernel_bandwidth} (Table \ref{tab:kernel_bandwidth}).

\subsection{One-Vs-All Classifier} 
\label{sec:ovaclassifier}
The one-vs-all (OVA) classifier is one of the two multiclass classifiers evaluated in this study. The OVA classifier aggregates the results of the binary classifiers outlined in Section \ref{sec:binarymodel} to determine the relative probabilities among classes. Generally, OVA classifiers use the maximum assigned probability across a range of binary classifiers to assign a classification, but we extend this to calculate a probability. The binary probabilities for each class, $P_B(t_k = 1 | \textbf{x})$ in Equation \ref{eq:binarybayes}, are normalized in order to determine the comparative probability of each transient class:
\begin{equation}
    P_O (t_k=1|\textbf{x}) = \frac{P_B(t_k=1|\textbf{x})}{\sum_{k'=1}^{K} P_B(t_{k'}=1|\textbf{x})},
    \label{eq:ovaprob}
\end{equation}
where $K$ is the number of classes. This aggregation of binary classifiers constitutes a multiclass classifier, the results of which we may use to determine the relative likelihood among considered classes of transients. The comparative probabilities among classes allows us to classify transient events by selecting the class with the maximum probability assigned: 

\begin{equation}
\argmax_{k \in K} P_O(t_k=1|\textbf{x}),
\label{eq:ovaprediction}
\end{equation}
where $K$ is the set of all disjoint classes, as outlined in Figure \ref{fig:classhierarchy2}, and $P_O(t_k=1|\textbf{x})$ is the probability of the class $i$ given galaxy data $\textbf{x}$, given by Equation \ref{eq:ovaprob}.

\subsection{KDE Multiclass Classifier} 
\label{sec:multimodel}
We develop a multiclass classifier using multivariate kernel density estimates per class to directly compare the likelihood of each transient class to one another. Whereas the OVA classifier estimates the positive space and negative space for each class distinctly, resulting in  $2*K$ probability density distribution estimates, the KDE multiclass classifier as outlined here requires only $K$ kernel density estimates, where $K$ is the number of classes. This reduction in probability density estimation is expected to result in enhanced performance when compared to the OVA model.

We optimize these bandwidths and kernels using the validation data likelihood per class, rather than minimizing misclassification loss as in the binary classifiers. As noted by \citet{doi:10.1198/004017005000000391}, optimizing separate bandwidths for many classes using classification is expensive, because the roles of the bandwidths in classification are interdependent. Thus, a thorough search entails trying all possible combinations of the bandwidths, which is impractical. On the other hand, we can search for good bandwidths for each class independently using the likelihood of validation data. Assuming that our model is good, and we have sufficient data, this approach should give better overall probability estimates for class membership. For some scenarios, we may be giving up some classification performance compared to tuning the bandwidths for that task if it were feasible.

We use kernel density estimation to determine the probability density for the samples of each class separately. To find the likelihood for a sample $\textbf{x}$, we normalize over the probability densities of all classes. The resulting multiclass probability for a particular transient class $k$ and galaxy with features $\textbf{x}$ is thus:
\begin{equation}
P_M(t_k=1|\textbf{x}) = \frac{p(\textbf{x}| t_k = 1)}{ \sum_{k'}^K p(\textbf{x}| t_{k'} = 1)},
\label{eq:multimcprob}
\end{equation}
where the $p(\textbf{x}| t_k = 1)$ is the likelihood for the kernel density estimate for class $k$. As opposed to Equation \ref{eq:binarybayes}, which normalizes over the positive and negative class densities, this multiclass probability is computed by normalizing over the probability densities of each class. 

As with the OVA classifier, we classify a sample with the maximum probability class: 

\begin{equation}
\argmax_{k \in K} P_M(t_k=1|\textbf{x}),
\label{eq:prediction}
\end{equation}
where $K$ is the set of all disjoint classes, as outlined in Figure \ref{fig:classhierarchy2}, and $ P_M(t_k=1|\textbf{x})$ is the multiclass probability of the class $k$ given galaxy data $\textbf{x}$ (Equation \ref{eq:multimcprob}).

 \subsection{Preprocessing}
 
We use 19 features---including 10 photometric bands and nine derived colors---in our models.
Given the range of values among the colors versus original host galaxy magnitudes, we apply a standard scaling to ensure that the range of KDE bandwidths considered will be appropriate for the data. We scale by deducting the mean and scaling to unit variance for each feature independently:
\begin{equation}
    \hat{x_i} = \frac{x_i-\mu}{\sigma},
    \label{eq:scaling}
\end{equation}
where $\hat{x_i}$ represents the scaled feature value for datapoint $i$ for a particular feature,  $x_i$ is the original value, and $\mu$ and $\sigma$ are the  mean and standard deviation of the training data values for that feature: $\sigma=(\sum\left(x_{i}-\mu\right)^{2}/N )^ {1/2}$. 

Our magnitudes and colors are measured in the observer’s frame. While rest-frame colors are more closely linked to the stellar populations of galaxies and hence supernova types, $k$-correcting the observed data would depend on photometric redshifts and introduce other biases and errors. Thus, we postpone the exploration of how the photometric redshifts generated by LSST and the $k$-corrections derived from them would affect our analysis until future work.

\subsection{Evaluation Strategy}
\label{sec:evaluation}
We aggregate the performance for our classifiers using 10-fold cross-validation, which provides adequate data support for training small classes (our smallest class, Ia-HV, has 42  host galaxies). Here, we cycle over the folds, using each 10\% of the data for testing and the remaining 90\% for training. We optimize the bandwidth for each class by further splitting the training data into $70\%$ training and $30\%$ validation, and evaluating the bandwidth on the validation set.  We compute the average purity and completeness per class over the $10$ test folds for each classifier.

\subsubsection{Performance Measures}
\label{sec:measurements}

The balanced purity (Equation \ref{eq:balancedpurity}) is equivalent to purity (also referred to as precision)  under the condition that all classes are represented equally in the test set. To compute balanced purity, we weight the number of true positives (TP) and false positives (FP) by the number of samples of the corresponding class.  The resulting balanced purity is 
\begin{equation}
    \textrm{BalPurity}(t_k) = \frac{TPR}{ TPR + \sum_{k', k' \neq k}^K \frac{FP_{k'}}{count(k')}},
    \label{eq:balancedpurity}
\end{equation}
where $TPR= TP_k/ count(k)$ and $FP_{k'}$ is the number of samples predicted as class $k$ but are actually of class $k'$, and $K$ represents the entire set of classes.

Balanced purity may be interpreted as the purity  our model would have if all classes in the test set were equal sizes. It allows for more interpretable aggregated measurements. Otherwise, the classes would have baselines based on their class size, and their relative performances might be difficult to compare to each other and to their random baselines (\S \ref{sec:baselines}). In cases where purity (not balanced purity) is used, as in the evaluation of uniform versus frequency-based priors in Section \ref{sec:incpriors}, we follow the standard definition:
\begin{equation}
    \textrm{Purity}(t_k) = \frac{TP}{TP + FP}.
    \label{eq:regpurity}
\end{equation}

The completeness is a measure of the percentage of events per class that the classifier is able to accurately identify (also referred to as recall, true positive rate, or specificity):
\begin{equation}
    \textrm{Completeness}(t_k) = \frac{TP}{TP + FN}.
    \label{eq:completeness}
\end{equation}

The balanced purity and completeness per class averaged over 10 folds are visualized in Figures \ref{fig:binary_g_W2_metrics} and \ref{fig:multi_g_W2_metrics} for the binary and multiclass classifiers, respectively. The significance of these measures is denoted by the confidence intervals discussed in Section \ref{sec:intervals}. The dotted red lines signify the expected random baselines, as described in Section \ref{sec:baselines}.

\subsubsection{Confidence Intervals}
\label{sec:intervals}
\begin{figure*}
    \centering 
    \includegraphics[width=0.8\textwidth]{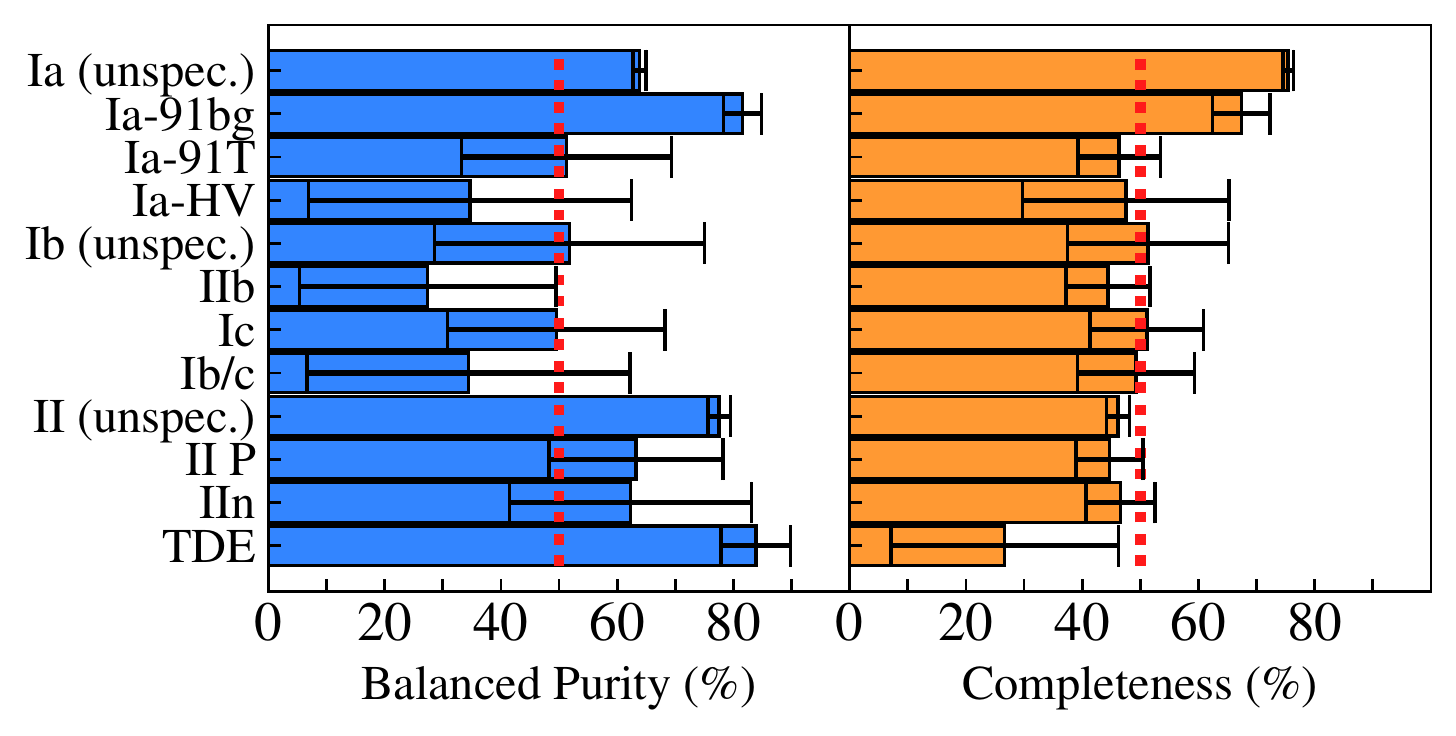} 
    \caption{
    For binary classifiers trained on THEx dataset using 10-fold cross validation: the balanced purity and completeness for each class annotated with 95\% confidence intervals and random baselines (described in Section \ref{sec:evaluation}).  Ia (unspec.),  Ia-91bg,   II (unspec.), and TDE are identified with balanced purity rates above random.  Ia-91bg and TDE perform most significantly above random in terms of balanced purity, exceeding it by $32-34\%$. Ia (unspec.) and Ia-91bg also perform above-random with respect to completeness. }
    \label{fig:binary_g_W2_metrics}
\end{figure*}

When finding the average balanced purity and completeness across $k$-folds (Figures \ref{fig:binary_g_W2_metrics} and \ref{fig:multi_g_W2_metrics}), we take into account the 95\% confidence intervals. We assume the balanced purity and completeness per fold are Gaussian distributed. 
The standard deviation of the performance of the testing folds is calculated as:
\begin{equation}
    \sigma = \sqrt{\frac{1}{D-1} \sum_{i=1}^D (m_i -\mu)^2},
    \label{eq:std}
\end{equation}
where $D$ is the number of folds in $k$-fold cross validation and $m_i$ is the measurement being computed, either balanced purity (Equation \ref{eq:balancedpurity}) or completeness over each fold. 

The standard error of the mean is then calculated as:
\begin{equation}
   SEM = \frac{\sigma}{\sqrt{D} }.
\end{equation}
The resulting confidence intervals for the 95\% percentile are  $\mu + 1.96SEM, \mu - 1.96SEM$. These intervals represent how well our model performs overall. Meaning, if we construct a new set of $D$-folds for cross validation and evaluate our methods on the new splits, we expect  the new performance measures (average balanced purity over 10 folds and average completeness over 10 folds) to fall within these ranges.

\subsubsection{Random Baselines}
\label{sec:baselines}
We compute a random baseline per class to compare the results to, by considering a theoretical classifier that assigns class labels randomly. Because we focus on a likelihood-only model, the corresponding random baseline classifier does not incorporate prior knowledge of class frequency and randomly predicts classes at uniform rates. The random baseline for balanced purity (Equation \ref{eq:balancedpurity}) is based on the number of classes (as all classes are predicted uniformly randomly):
\begin{equation}
    \textrm{BalPurityBaseline}(t_k) = \frac{1}{K},
    \label{eq:puritybasline}
\end{equation} 
where $K$ is the total number of classes. In the binary case, $K=2$; in the multiclass case, $K=12$.  The random baseline for completeness is the same:
\begin{equation}
     \textrm{CompBaseline}(t_k)  = \frac{1}{K}.
    \label{eq:compbasline}
\end{equation}


\section{Experimental Results} 
\label{sec:results}

We evaluate the binary classifiers, the OVA classifier, and KDE multiclass classifier using the strategy outlined in Section \ref{sec:evaluation}.  
We measured the time taken to train and test our classifiers on 11260 data sources on an 11th Gen Intel core i7-11700 @2.5 GHz CPU using 12 cores. It took about 15 minutes per fold in 10-fold cross validation for the binary and OVA classifiers and less than one minute for the KDE multiclass classifier.

We consider several guiding questions in this analysis:

\begin{enumerate}
    \item How well can any one class be distinguished from all the rest of the classes using host galaxy photometric data alone? (\S \ref{sec:binary_results}, binary classification) 
    \item How well can we determine the most likely class among a range of transient classes using host galaxy photometric data alone? (\S \ref{sec:multi_results}, multiclass classification)
    \item Is it beneficial to use probabilistic estimation to improve purity for a subset of host galaxies? (\S \ref{sec:prob_performance})
    \item How does the performance of our models compare to alternative approaches, which either use other host galaxy features or the light curves of transients? (\S \ref{sec:light_curves})
    \item Is our model, and the corresponding performance measures, directly applicable to LSST? (\S \ref{sec:Zsection})
\end{enumerate}

\begin{figure*}
    \centering
    \includegraphics[width=0.95\textwidth]{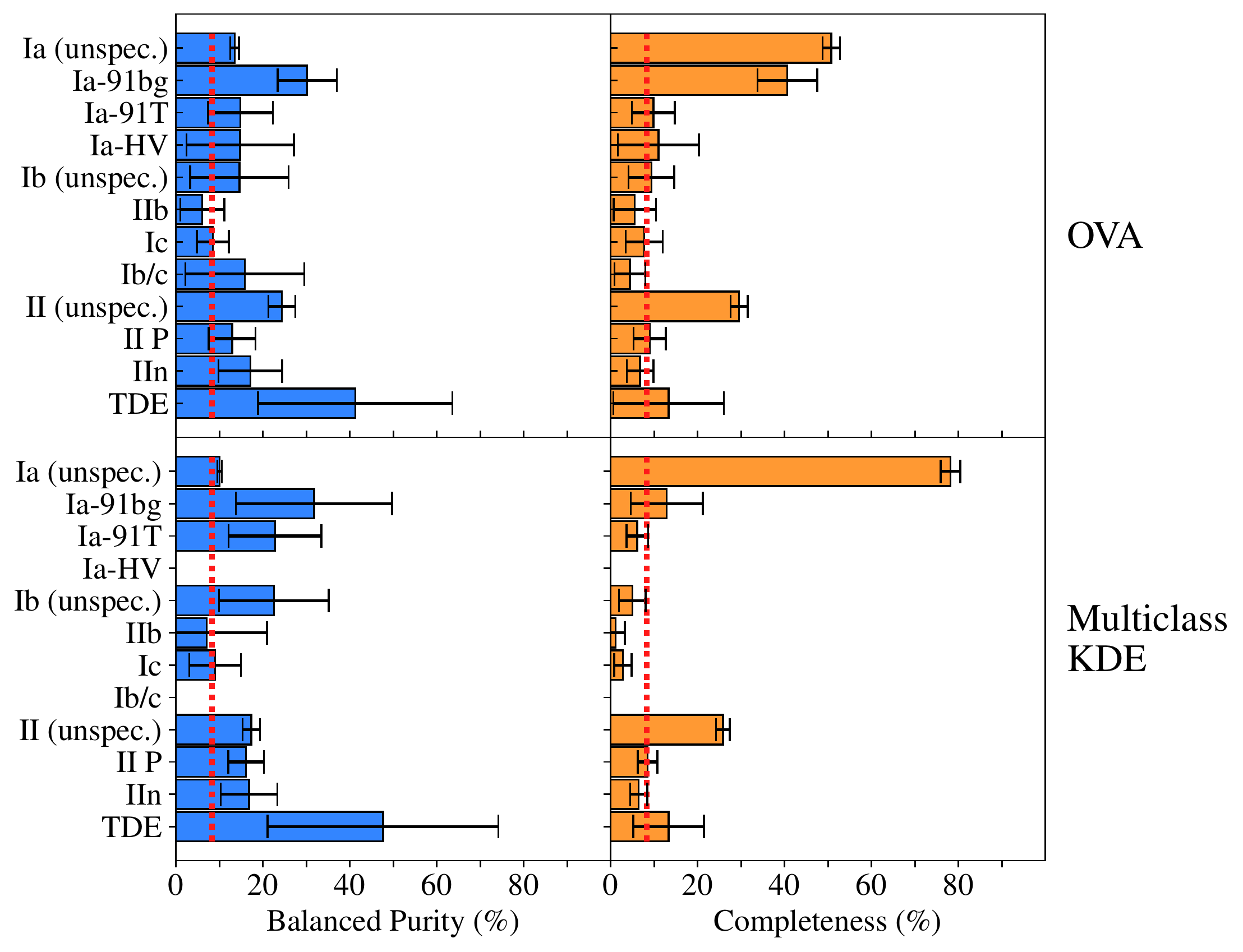}
    \caption{
     The balanced purity and completeness for the OVA classifier (top row) and KDE multiclass classifier (bottom row), annotated with $95\%$ confidence intervals and random baselines (described in Section \ref{sec:evaluation}). For both classifiers, above-random balanced purity is achieved for Ia (unspec.), Ia-91bg, II (unspec.), IIn, and TDE. OVA has one advantage in that it achieves higher completeness for Ia-91bg. The KDE multiclass classifier achieves higher completeness for Ia (unspec.) and higher purity for Ia-91T, Ib (unspec.), and II P.
     Ia-HV, IIb, Ic, and Ib/c all do not achieve above-random purity in any model, which may be due to insufficient data (e.g., 42 samples for Ia-HV and 92 samples for Ib/c) or that the host galaxies are not photometrically distinct.
     The purity and completeness in multiclass classification are zero for Ia-HV and Ib/c, because the training model predicts zero true positive cases for the test set.
     Overall, the KDE multiclass classifier performs better on these average balanced purity measures as well as in its probability performance, as discussed in Section \ref{sec:prob_performance}.  }
    \label{fig:multi_g_W2_metrics}
\end{figure*}

\subsection{Identifying Transient Classes Independently Using Binary Classifiers}
\label{sec:binary_results}
Figure \ref{fig:binary_g_W2_metrics} shows the balanced purity and completeness of each class, based on the performance of that class's binary classifier across the 10 test sets from 10-fold cross validation. A class that attains a balanced purity or completeness above the random baseline, and within a 95\% confidence interval that does not overlap the baseline, is considered to have achieved significant performance. We achieve this purity for four of the 12 classes:
Ia (unspec.) ($64\%\pm1\%$, where $\pm$ indicates the 95\% confidence limit),
Ia-91bg ($82\%\pm3\%$),
II (unspec.) ($78\%\pm2\%$), 
and TDE ($84\%\pm6\%$).
Ia-91bg, II (unspec.), and TDE stand out as exceptionally well-performing classes that outperform the random baseline for balanced purity by $28-34\%$. Ia (unspec.) and Ia-91bg also perform $17-25\%$ above random completeness.

\subsection{Distinguishing Transient Classes From One Another Using Multiclass Classifiers}
\label{sec:multi_results}
Although the binary classifiers clearly demonstrate that certain transient classes may be uniquely distinguished from all other classes, they do not address the question of whether multiple transient classes are distinguishable from one another and to what degree. The performances of the OVA classifier (which normalizes over the binary classifiers) and the KDE multiclass classifier indicate how well we may distinguish between different classes of transients at once (Figure \ref{fig:multi_g_W2_metrics}).

The KDE multiclass classifier achieves above-random ($>8\%$) balanced purity for eight classes:
Ia (unspec.) ($10\%\pm1\%$\, where $\pm$ indicates the 95\% confidence limit),
Ia-91bg ($32\%\pm18\%$),
Ia-91T ($23\%\pm11\%$),
Ib (unspec.) ($23\%\pm13\%$),
II (unspec.) ($17\%\pm2\%$),
II P ($16\%\pm4\%$),
IIn ($17\%\pm6\%$),
and TDE ($48\%\pm27\%$).
Ia-91bg and TDE are most significantly above the random baseline for balanced purity (as was the case for the binary classifiers), exceeding it by  23\%-39\%. Ia (unspec.) and II (unspec.) perform $69\%$ and $17\%$ above random completeness, respectively.

\begin{figure*}[ht!]
    \centering   
    \includegraphics[clip, width=1\textwidth] {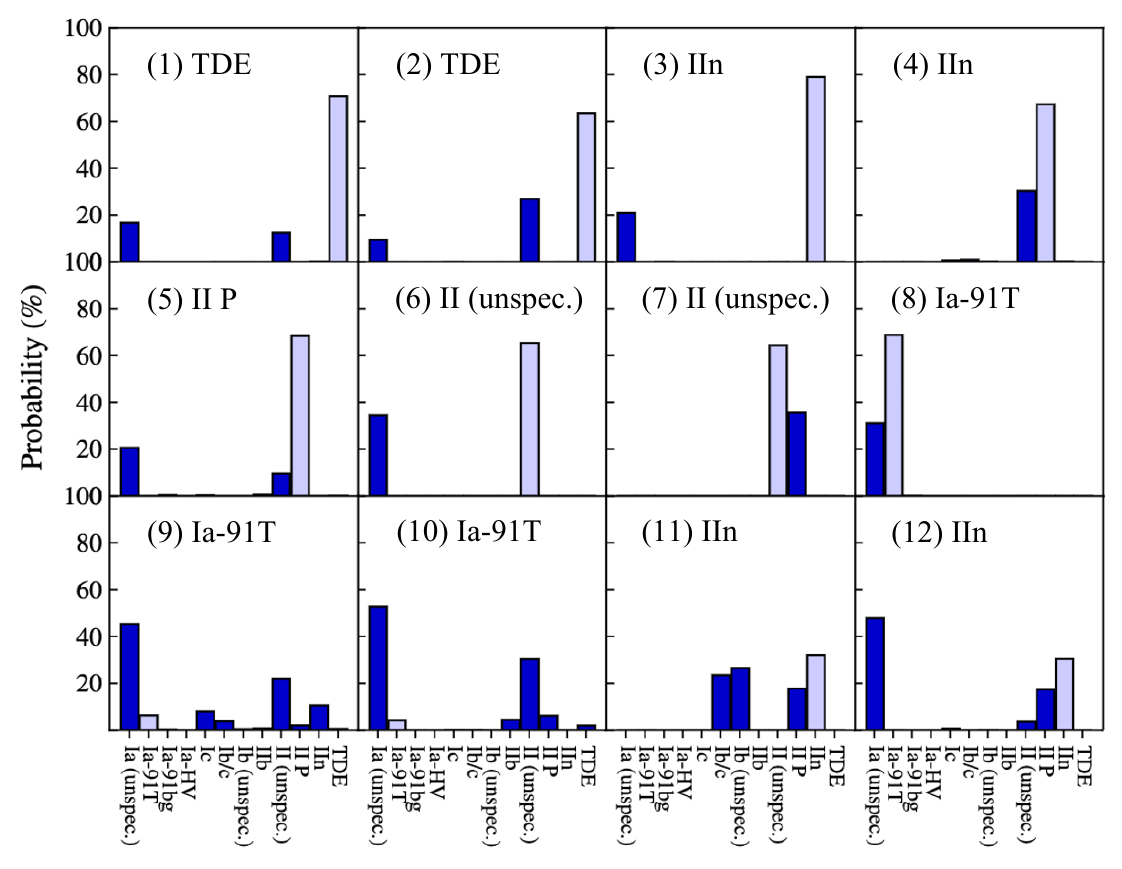}
    \caption{Example outputs for 12 different host galaxies from the KDE multiclass classifier run on all classes with uniform priors, where the true class is marked by the light purple bar and by the label beside the index number in each panel. These were selected from our dataset of 11260 host-transient pairs based on their demonstration of accurate prediction and/or utility of probabilities ($P_M$) provided by the model. Examples 1-8 show correct predictions, where the highest probability is assigned to the true class, for TDE (1,2), IIn (3), II P (4,5), II (unspec.) (6,7), and Ia-91T (8). Examples 9-12 demonstrate the utility of the probability distribution as a whole when considering the prediction. These events are  difficult to distinguish, and the classifier is not confident in any class. Taking the maximum probability prediction at face value would be erroneous, since a visual examination of the probability distribution reveals that no probability is $>50\%$, and there is some weight given to roughly half of the classes. These examples illustrate the utility of quantifying uncertainty and demonstrate that we are able to identify those host-transient pairs that are difficult to predict. 
    }
    \label{fig:exampleoutputs}
\end{figure*}

Figure \ref{fig:exampleoutputs} shows example outputs for 12 host galaxies evaluated by the KDE multiclass classifier (via Equation~\ref{eq:multimcprob}). These illustrate a range of valuable scenarios including both accurate transient class predictions (the maximum probability is associated with the true class; examples 1-8) and clear inaccurate predictions (which can be identified as unreliable, because the probabilities across all classes are low; examples 9-12).

OVA performs similarly, except that it does not achieve above-random purity for Ia-91T, Ib (unspec.), or II P, and it has significantly lower completeness for Ia (unspec.). OVA's main advantage is the above-random completeness for Ia-91bg ($41\%\pm7\%$). Ia (unspec.) and II (unspec.)  both achieve above-random balanced purity under both methods. The highest completeness is for Ia (unspec.) ($78\%\pm2\%$ for KDE multiclass classifier and $51\%\pm2\%$ for OVA), although the corresponding balanced purity narrowly out-performs the baseline. The majority of the rare classes which achieve above-random balanced purity are below-random with respect to completeness. This is the case for Ia-91bg, Ia-91T, Ib (unspec.), II P, IIn, and TDE for the KDE multiclass classifier and for IIn and TDE for OVA.

Overall, we observe that the KDE multiclass classifier achieves a higher balanced purity for more classes than OVA, and only performs worse for the completeness of Ia-91bg. There is also a distinct difference between these two classifiers in their variability in performance based on probabilities assigned to events, discussed in detail in the next section.

\subsection{Using Probabilities to Achieve Higher Rates of Purity}
\label{sec:prob_performance}

Can we improve on the balanced purity by considering only higher probability assignments? To test this, we evaluate how the balanced purity and completeness change as a function of assigned probability ($P \equiv P_B, P_O,$ or $P_M$, where $P_B$, $P_O$, and $P_M$ correspond to binary, OVA, and KDE multiclass classifiers in Equations \ref{eq:binarybayes}, \ref{eq:ovaprob}, and \ref{eq:multimcprob}, respectively).
For each transient class, we calculate the balanced purity and completeness for samples given a probability equal to or greater than a range of thresholds, increasing at $10\%$ intervals. For the LSST use case, we focus on those classes that attain a significant balanced purity with a non-negligible completeness. Because relatively few events can be followed-up, purity must be high to avoid wasting telescope resources, while completeness can be low.

\begin{figure*}
    \centering
     \includegraphics[width=0.88\textwidth]{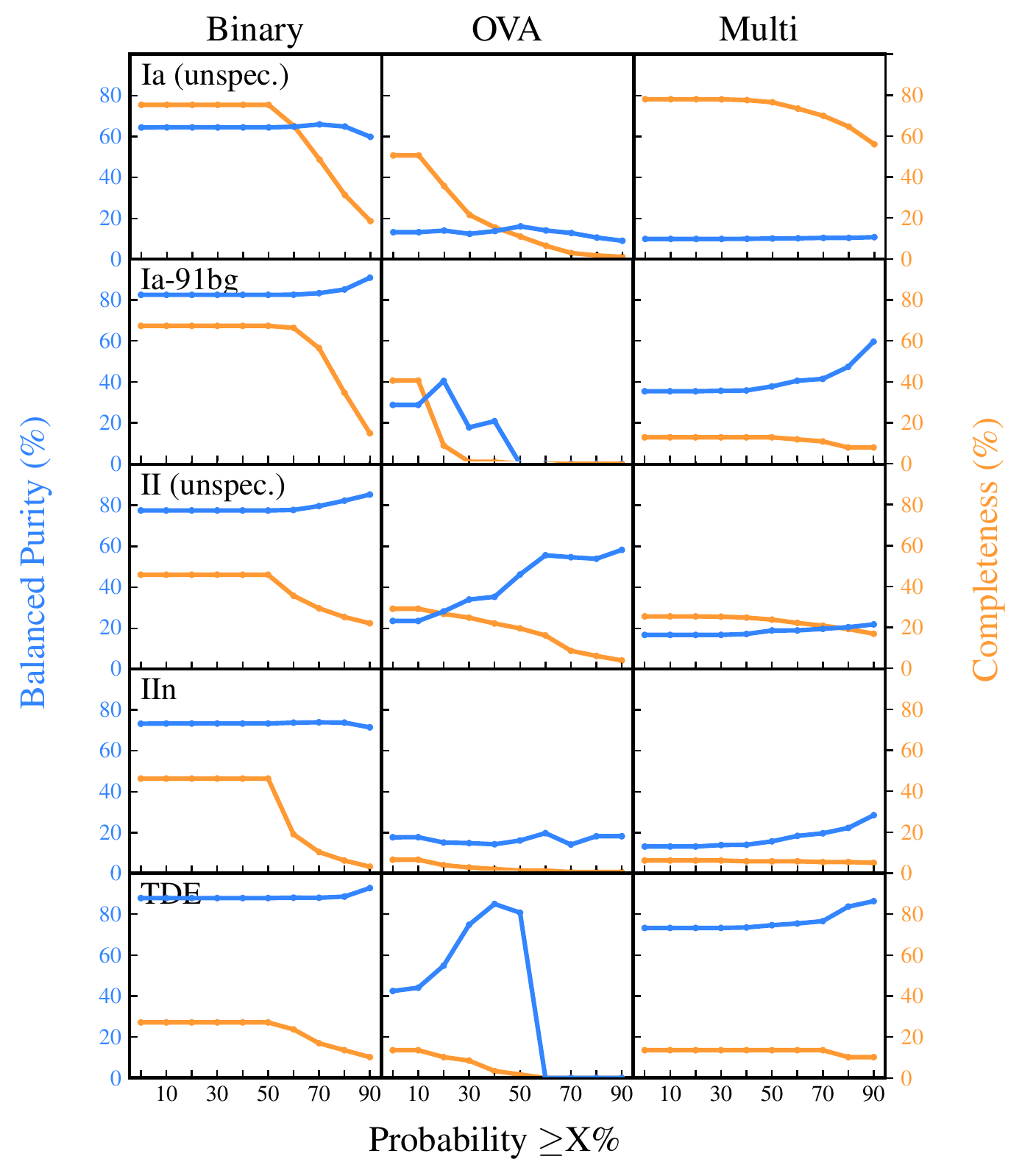}
    \caption{ Examples of balanced purity (blue) and completeness (orange) as a function of assigned probability threshold (see Appendix \ref{sec:appendix_classrates} for details).
    The threshold intervals are each stepped by 10\%.
    For these five transient classes, the balanced purity significantly improves at higher probability thresholds for the multiclass classifier (third column).
    These curves are based on all the test data, considered all at once (although divided into probability ranges). This is somewhat different than the averages computed in Figures \ref{fig:binary_g_W2_metrics} and \ref{fig:multi_g_W2_metrics}, which consider the variance of performance across each 10 fold of cross validation, and results in somewhat different measures for $P \geq 0\%$.
    }
    \label{fig:pr_curves}
\end{figure*}

Figure \ref{fig:pr_curves} shows the balanced purity-completeness curves for five particularly well-performing transient classes: Ia (unspec.), Ia-91bg, II (unspec.), IIn, and TDE.  Here the balanced purity significantly improves for the KDE multiclass classifier between $P_M\geq0\%$ and 90\%. Overall, the classifiers often identify classes at higher rates of balanced purity and lower completeness as the probability threshold increases. This is in line with the understood relationship between the two: as the accuracy of identification improves, we are able to identify fewer events in a class at the enhanced rate of purity. Appendix \ref{sec:appendix_classrates} details how the curves are calculated and shows an expanded version of Figure \ref{fig:pr_curves} for all classes (Figure \ref{fig:pc_merged_all}).

\subsection{Comparison to Alternative Classification Approaches} 
\label{sec:light_curves}

The dataset in this study is novel in its breadth of transient classes and use of host galaxy photometric magnitudes. There are no perfectly comparable methods to consider, because other approaches generally focus on only a few classes, often Type Ia and core-collapse (CC) supernovae, and use a different set of features (incorporating transient information or galaxy features that are not easily obtained for most LSST-observed galaxies).  

The most similar approach to our own is that of \citet{Gagliano:2020ucg}, which uses a random forest classifier to distinguish Ia and CC using 317 host galaxy features from Pan-STARRS and seven features regarding the transient event. They use 5-fold cross validation and rebalance the training data for each fold to ensure that Ia and CC each have 3500 samples in the training set (by sampling 609 CC from a set of 2891 and sub-sampling Ia down to 3500). This balanced training set ensures that the random forest classifier does not use class frequency information, an equivalent assumption to our uniform priors. The SNe Ia events in their dataset correspond to our Ia (unspec.), and their core collapse set consists of II, IIb, II P, and Ib/c. Considering the transient class hierarchy (Figure \ref{fig:classhierarchy2}), classes not belonging to the parent class Ia belong to CC, except for TDE. In our dataset, TDE make up only 59 host galaxies, compared to 11201 Ia and CC host galaxies (the remainder of the dataset). For this reason, we may compare the performance of our  binary Ia (unspec.) classifier to that of the \citet{Gagliano:2020ucg} Ia versus CC  classifier. Both attain a balanced purity of 64\% for SNe Ia, despite the fact that we use only 19 host galaxy features (as opposed to their 317) and no transient information (as opposed to their seven features regarding the transient). We furthermore verify that we may achieve similar performance on LSST data, at least for Ia (unspec.) and II (unspec.) (\S \ref{sec:zbias}).

In addition to Ia versus CC, we may consider the other findings of \citet{Gagliano:2020ucg} regarding rarer classes. \citet{Gagliano:2020ucg} were unable to construct a classifier that could distinguish between SLSNe, SNe II P, SNe IIb, SNe IIn, and SNe Ib/c, which they attributed to a lack of data. However, they did observe a distinction in the distribution of host galaxy features for SLSNe, SNe II P, and SNe IIb when projecting the data into a three-dimensional space using t-SNE \citep{vanDerMaaten2008}.
We do not classify SLSNe, but of the remaining rare classes, we are able to distinguish II P with above-random balanced purity with the KDE multiclass classifier ($16\%\pm4\%$). 
After applying a cutoff in assigned probability (\S \ref{sec:prob_performance} and Appendix \ref{sec:appendix_classrates}), we generally attain even higher rates of balanced purity (e.g., for IIn).

Overall, the methods presented here illustrate the utility of using host galaxy features in classifying transients, as compared to alternative approaches that use different host galaxy features. We attain similar performance for Ia (unspec.) with fewer features and better performance for some rare classes, notably IIn.
   
\begin{figure*}
   \centering
    \includegraphics[width=0.99\textwidth]{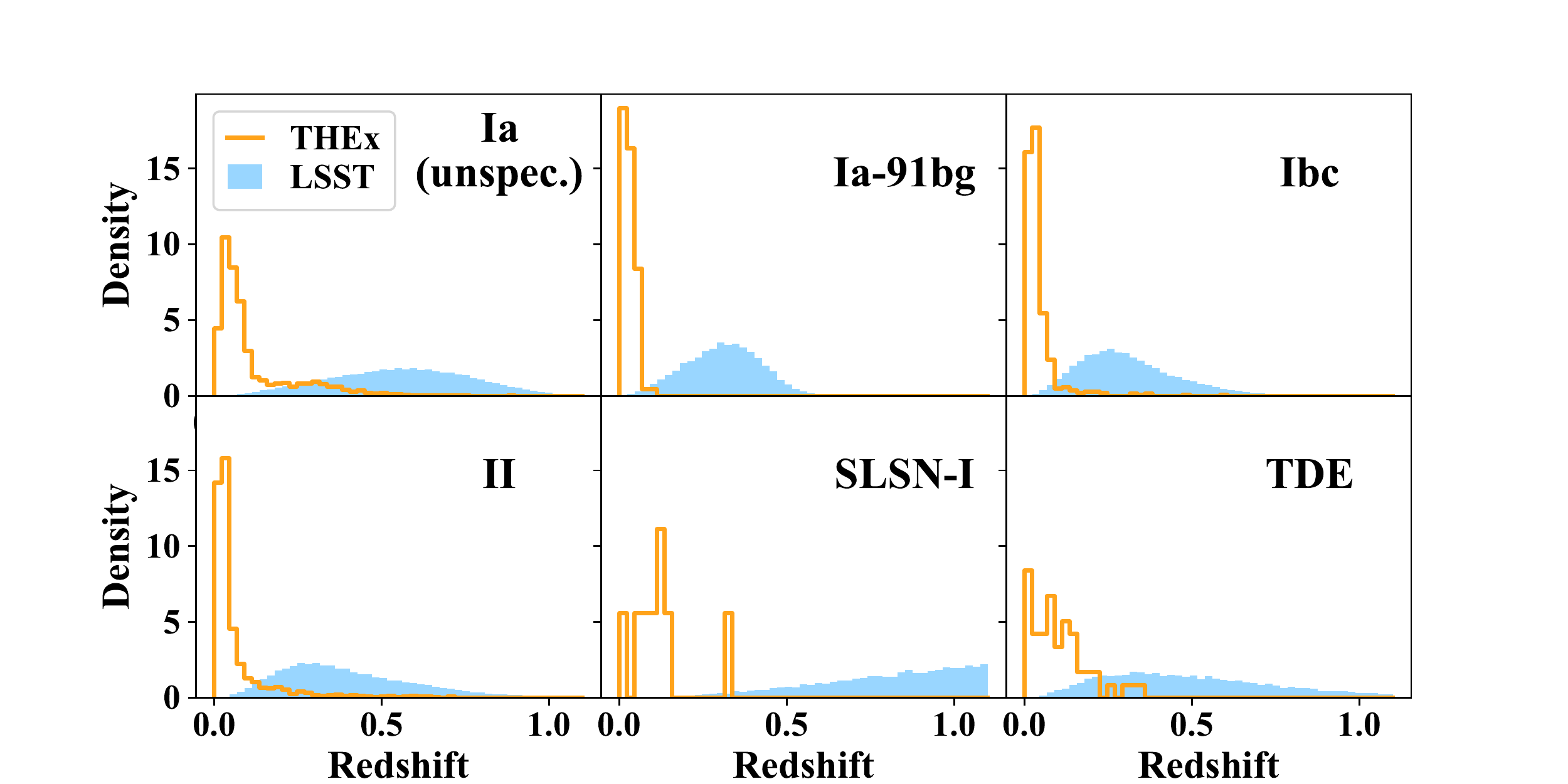}
    \caption{The redshift distributions for  Ia (unspec.), Ia-91bg, Ibc, II, SLSN-I, and TDE in our dataset versus the anticipated LSST data. The LSST data is approximated by the Photometric LSST Astronomical Time-series Classification Challenge (PLAsTiCC) dataset. The PLAsTiCC data is Gaussian distributed and centers at a redshift $0.3-0.6$ higher than our data. These varying distributions warrant an investigation into determining the potential biases our model may have towards  differently distributed data (\S \ref{sec:Zsection}). }
    \label{fig:zdistributionsgW2}
\end{figure*}


\begin{figure}
   \centering
    \includegraphics[width=\thirdimgwidth\textwidth]{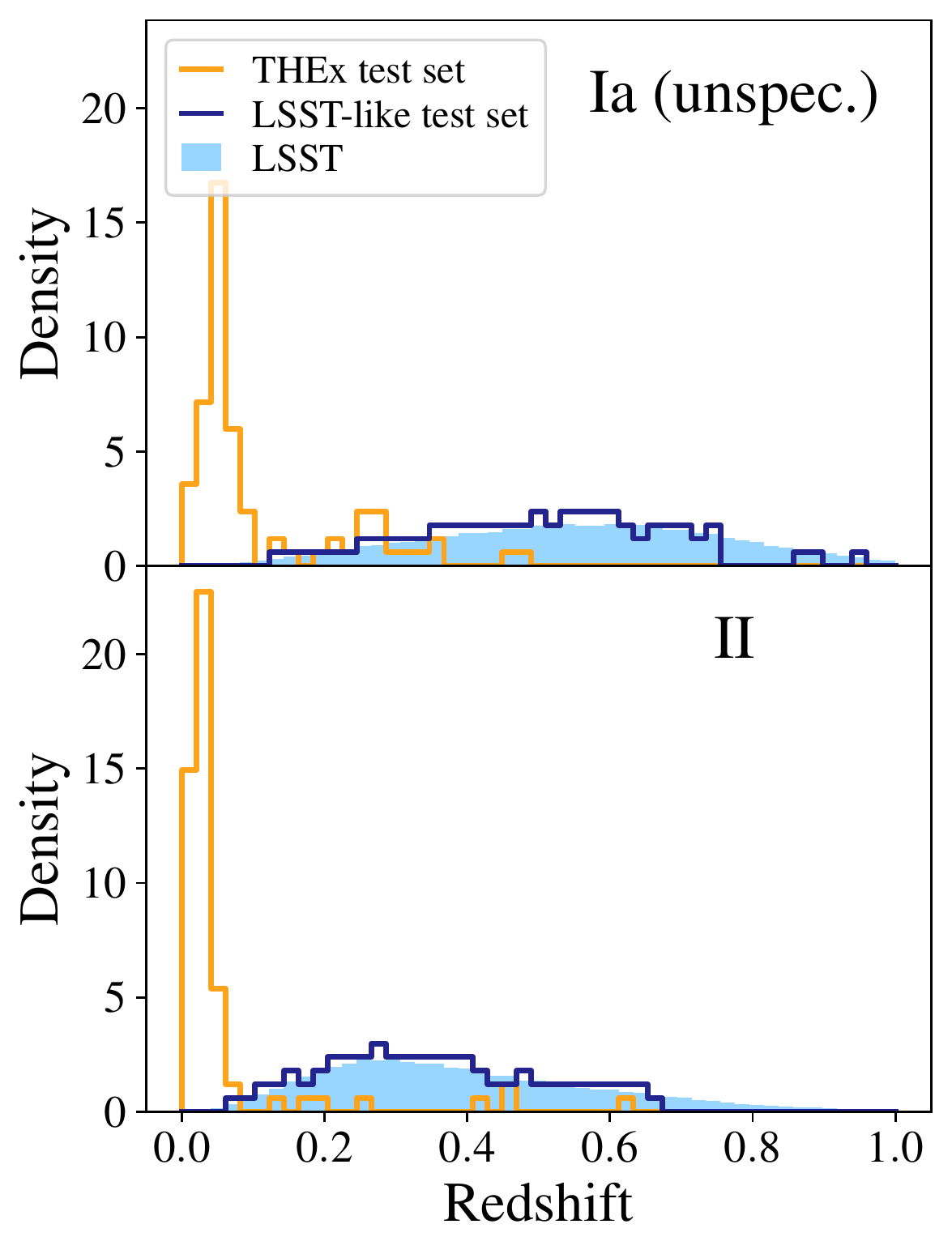}
    \caption{ Redshift distributions of Ia (unspec.) and II in the THEx dataset sampled with the same distribution as LSST-like PLAsTiCC data (\textit{LSST-like test set}) versus randomly sampled (\textit{THEx test set}). The expected distribution for LSST (light blue distribution in the background) is based on the PLAsTiCC dataset and is the same as in Figure \ref{fig:zdistributionsgW2}. We compare the performance of our model on these differently distributed test sets to evaluate redshift bias in Section \ref{sec:zbias}. The two test sets visualized here are from one of the 10 trials.   }
  \label{fig:zdistssampled}
\end{figure}

\begin{figure*}
   \centering 
    \includegraphics[width=0.8\textwidth]{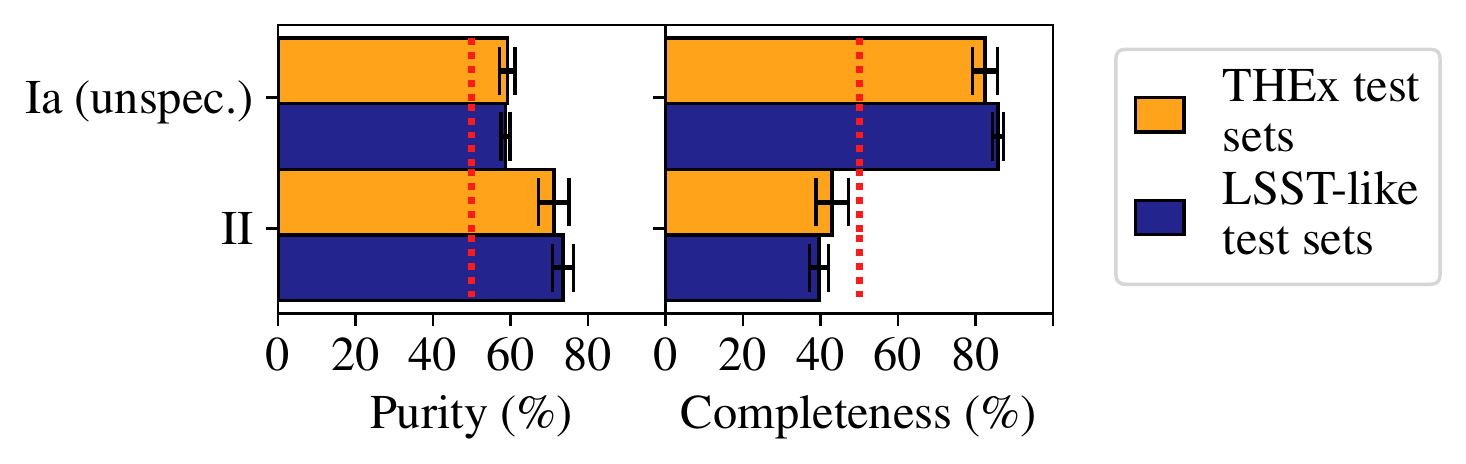} 
    \caption{The purity and completeness of Ia (unspec.) and II  on \textit{THEx test set}  versus \textit{LSST redshift distribution-like test set},  averaged over 10 trials of the KDE multiclass classifier on test sets with equal numbers of each class. The performance on the two test sets is roughly the same within the 95\% confidence intervals. For both test sets, the purity of Ia (unspec.) is $59\%$ and the purity of II is $71-74\%$;  the completeness of Ia (unspec.) is $82-86\%$ and for II is $40-43\%$. Because there is no distinction between the performance of these differently distributed test sets,  we conclude our model is readily applicable to LSST data for SNe Ia and II.  Note that here we use the standard purity measure, $TP/(TP+FP)$.}
     \label{fig:ztest}
\end{figure*}

\subsection{Applicability to LSST data}
\label{sec:Zsection}

In this section, we investigate the applicability of our methods to LSST. We address the two notable differences between our data and LSST: the distribution of each class's data over redshift and the relative frequency of classes. We examine our dataset relative to events anticipated by LSST and adjust our dataset accordingly (\S \ref{sec:lsstclasses}). With the adjusted data, we evaluate the impact of the differing redshift distributions by ensuring the classification and modeling biases due to redshift are consistent between our data and the anticipated LSST data (\S \ref{sec:zbias}). The difference in class frequencies between our data and LSST has been implicitly handled in previous sections, because we use a likelihood-only model and report balanced purity (rather than standard purity, which is based on relative class size). However, in order to estimate the performance of our model on LSST, we consider the anticipated rates of transient classes detected by LSST and the performance of our model when  using similar rates as priors (\S \ref{sec:incpriors}). We use the resulting performance measurements to estimate the number of transient events detected by LSST for which our approach may enable immediate follow-up (\S \ref{sec:anticipatedrates}).  
 
\subsubsection{Adjusting Classes for LSST-like Tests}
\label{sec:lsstclasses}

We estimate the distribution and classes of LSST data using the test dataset from the Photometric LSST Astronomical Time-series Classification Challenge (PLAsTiCC). Their test dataset consists of 19 classes to be observed by LSST and simulates their corresponding rates \citep{2018arXiv181000001T, Kessler_2019}. However, because initial observation will determine whether these events are likely extra-galactic, we need only consider the extra-galactic classes, of which there are nine in the training set and three additional in the test set. 

The extra-galactic transient classes in PLAsTiCC are SN Ia, SN Ia-91bg, SN Iax, SN II, SN Ibc, SLSN-I, TDE, Kilonova (KN), Active galactic nuclei (AGN), and three rare classes (ILOT, CaRT, and PISN). For the classes SN Ia, SN Ia-91bg, SLSN-I, and TDE, we use our classes as they are. We include SLSN-I despite it having only six samples, because it allows us to better simulate the performance of our model when applied to real data, which initially will have classes without adequate training data.  We construct class Ibc to mimic that of SN Ibc by combining our Ib, Ic, and Ib/c samples. We use SN II instead of II (unspec.), because the PLAsTiCC test set does not make a distinction of different Type II subclasses. Our II class consists of all II subclasses shown in Figure \ref{fig:classhierarchy2}. We exclude SN Iax, KN, AGN, and the three rare classes, because we have no data for them. The six classes we do consider make up 94\% of the extra-galactic transient data in the PLAsTiCC test set. 

\subsubsection{Evaluation of Redshift Bias in Training Data vs. LSST}
\label{sec:zbias}

LSST will observe more events farther away than previous optical surveys, making the distribution of the observed events over redshift different from the historical surveys on which our model is trained. This difference in distribution is visualized in Figure \ref{fig:zdistributionsgW2}, which shows the comparative redshift distributions for Ia (unspec.), Ia-91bg, Ibc, II, SLSN-I, and TDE in our dataset versus the anticipated LSST data (from the PLAsTiCC test set). 

Given the significant differences in the class distributions over redshift for these six classes (Figure \ref{fig:zdistributionsgW2}), we consider how the performance of our model may change when evaluated on data that is distributed more like LSST data. We determine how the performance of the model changes based on the redshift distribution of the test data by using two test datasets that differ only in their distributions over redshift.  We conduct this evaluation on only Ia (unspec.) and II, because these classes are prevalent enough to retain adequate data support across the entire sampled distribution.  For each test set, we sample each class from our dataset with redshift distributions consistent with those of the PLAsTiCC dataset to attain our \textit{LSST-like test set}. For the second test set, we randomly sample from our data to create the comparative \textit{THEx test set}. Each test set contains 82 Ia (unspec.) and 82 II. The distributions of an example test set are shown in Figure \ref{fig:zdistssampled}, which illustrates the \textit{LSST-like test set}  roughly covering the LSST distribution, although there is some loss of data at  high redshift, particularly for Ia (unspec.).

We conduct this analysis using the KDE multiclass classifier, outlined in Section \ref{sec:multimodel}. The experiment is repeated for 10 trials, and the resulting average purity and completeness per class are visualized in Figure \ref{fig:ztest}. For each trial, we sample each test set (as outlined above) and remove the test data from the training set. 

Figure \ref{fig:ztest} shows that the average performance on the \textit{LSST-like test set} is not much different than that on \textit{THEx test set}.  There are no dramatic differences in purity for the LSST-like test sets versus THEx test sets: $59\%\pm1\%$ versus $59\%\pm2\%$ for Ia and $74\%\pm3\%$ versus $71\%\pm4\%$ for II. Similarly for completeness: $86\%\pm1\%$ vs $82\%\pm3\%$ for Ia and $40\%\pm2\%$ versus $43\%\pm4\%$ for II. 

We conclude the proposed model is readily applicable to LSST data for Ia (unspec.) and II. In practice, this finding is only relevant for early classification of LSST alerts, when only historical data is accessible for training. As our dataset is expanded using newly observed and classified transients from LSST, the distribution over redshift in our dataset will more strongly resemble LSST-gathered data, avoiding the discrepancy in redshift distribution. This will allow us to  ensure the applicability across classes and redshift ranges, as well as likely improve classification accuracy across events.

 \begin{figure*}
\centering 
     \includegraphics[width=0.7\textwidth]{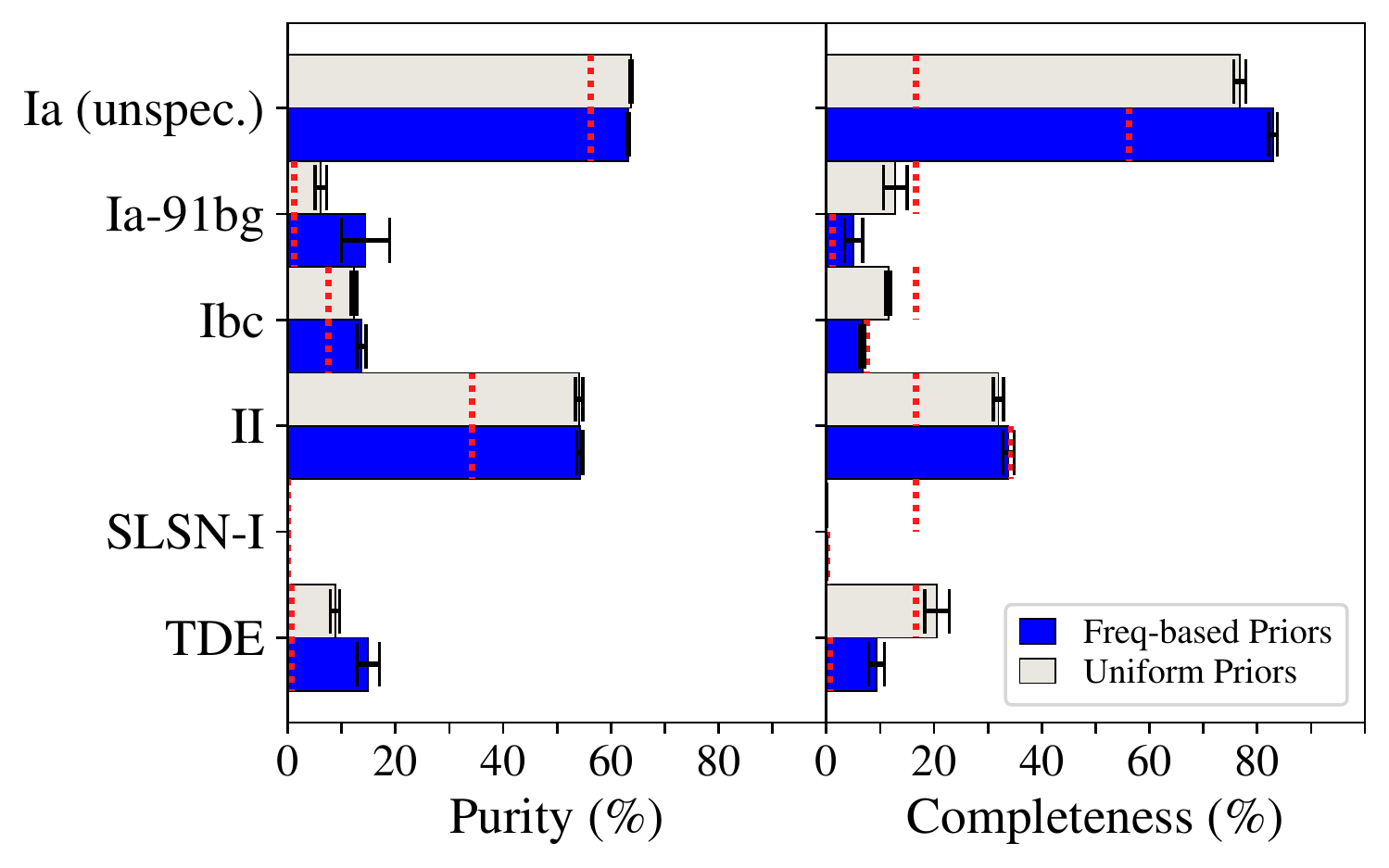} 
    \caption{ Standard purity ($TP/(TP+FP)$) and completeness of the KDE multiclass classifier with uniform priors versus frequency-based priors. These results are averaged over 10 repeated trials of 6-fold cross validation to ensure reliable estimates. The random baselines for completeness are different between the two, because the baselines for the frequency-based priors are based on a random classifier that uses the class frequency in prediction (Equation \ref{eq:wpriorsbaseline}), whereas the random baselines for the uniform priors are based on a random classifier that does not (Equation \ref{eq:compbasline}). We achieve above-random purity for all classes under both sets of priors. There is a significant increase in purity for TDE and Ia-91bg, the rarest classes besides SLSN-I,  with frequency based priors. For completeness, we see an increase for Ia (unspec.) and decrease for TDE and Ia-91bg with priors (although both completeness measures for Ia-91bg are below the random baseline).
    }
      \label{fig:multi_g_W2_sim_metrics}
\end{figure*}

\subsubsection{Incorporating Class Frequencies}
\label{sec:incpriors}

Besides different redshift distributions, the second critical difference between the dataset used here and the data collected by LSST is the relative frequency of classes. We consider the performance of our model on data with relative frequencies like those expected of LSST.  We study the change in performance when incorporating frequency-based priors into our KDE multiclass likelihood (Equation \ref{eq:multimcprob}) to get a posterior probability. We consider any improvements to our model when using class priors and try to simulate the LSST-like data as best as possible.

In the case of frequency-based priors, we assume that the random baseline classifier assigns classes at rates of frequency consistent with the priors. In this case, the purity baselines with and without priors, and the completeness baselines with priors, are the proportion of samples in the class:

\begin{equation}
     \textrm{PriorsBaseline}(t_k)  = \frac{count(t_k)}{\sum_{k'=1}^K count(t_{k'})}.
    \label{eq:wpriorsbaseline}
\end{equation}

We try to estimate our performance on LSST data by ensuring that our class distributions are as similar to LSST as possible. This allows us to use the resulting purity and completeness measures (with priors, depicted in Figure \ref{fig:multi_g_W2_sim_metrics}) to estimate the number of LSST alerts that we will be able to accurately classify using our method (Table \ref{tab:estTPFP} in Section \ref{sec:anticipatedrates}). Since standard purity, $TP/(TP+FP)$ is dependent on the class frequency (demonstrated by the random baseline, Equation \ref{eq:wpriorsbaseline}), we need to ensure that our class frequencies reflect those of LSST. 
We randomly sample our largest class, Ia (unspec.), down to 4800 events, which adjusts its frequency prior to be roughly that of LSST (as shown in Table \ref{tab:priors}). The prior frequencies of the other THEx transient classes analyzed in this section are consistent with LSST to within 1-2\%.
We threshold the lowest prior to be a minimum of 0.1\% (i.e., the proportion of SLSN-I in the dataset is $<0.1\%$, but we use a $0.1\%$ as its prior probability).

\begin{table}[]
\centering  
\begin{tabular}{r|cr|cr}
\multicolumn{1}{c|}{\textbf{Class}}  & \multicolumn{2}{c|}{\textbf{LSST [$yr^{-1}$]}}                           & \multicolumn{2}{c}{\textbf{THEx}}    \\
\multicolumn{1}{c|}{ } & Count  & \multicolumn{1}{c|}{Prior} & Count  & \multicolumn{1}{c}{Prior} \\ 
\hline
Ia (unspec.) & \multicolumn{1}{r|}{553,277} & 57\%  & \multicolumn{1}{r|}{4,800*} & 56\%   \\
SNe Ia-91bg & \multicolumn{1}{r|}{13,398}  & 1\%   & \multicolumn{1}{r|}{101} & 1\%  \\
Ibc & \multicolumn{1}{r|}{58,365 } & 6\%  & \multicolumn{1}{r|}{643} & 8\%  \\
SNe II & \multicolumn{1}{r|}{333,383 }  & 34\%    & \multicolumn{1}{r|}{2,923 }    & 34\% \\ 
SLSN-I & \multicolumn{1}{r|}{11,927}  & $1\%$     & \multicolumn{1}{r|}{6}    &  $<1\%$ \\ 
TDE & \multicolumn{1}{r|}{4,518}  & $<1\%$     & \multicolumn{1}{r|}{57}    &  $1\%$ \\ 
\end{tabular}       
    \caption{Frequencies and corresponding priors  for six classes in our dataset for which we have known LSST rates. The expected LSST counts are based on the PLAsTiCC test dataset, and their priors are relative only to these six classes of extra-galactic transients \citep{Kessler_2019}. These classes consist of 94\% of all anticipated extragalactic transient alerts from LSST.  We use THEx priors in Section \ref{sec:incpriors} to compare performance of the model with and without priors. 
    }
    \label{tab:priors}
\begin{tablenotes}
\item[*] *Ia (unspec.) is subsampled down to 4,800 events in order to adjust its prior frequency to roughly that expected for LSST. All the other transient classes in our THEx dataset shown here are also consistent with those expected for LSST to within 1-2\%.
This allows us to reliably use the resulting purity estimates in Figure \ref{fig:multi_g_W2_sim_metrics} to estimate the number of accurate classifications that we can make on LSST data using our classifier (\S \ref{sec:anticipatedrates}).
\end{tablenotes}
\end{table}

Figure \ref{fig:multi_g_W2_sim_metrics} compares the average purity and completeness from 10 trials of 6-fold cross validation of the KDE multiclass classifier with and without incorporating frequency-based priors for our dataset. 
We use six folds here as it is easier to arrange relatively equal numbers of each rare class in each test and validation set. We further use repeated $k$-fold cross validation, because the average performance among the folds is subject to the variance in the division of folds and the sampling of the Ia class (down to 4800 samples). The repetition of the 6-fold cross validation accounts for this variance. We conduct 10 trials, where each trial randomly samples Ia and randomly divides the data into six folds. We report the resulting average and corresponding confidence intervals of those 10 measures (for purity and completeness), and use these measures to estimate the number of TP and FP predictions from LSST in Section \ref{sec:anticipatedrates}. 

With or without priors, Ia (unspec.), Ia-91bg, Ibc, II, and TDE achieve above-random purity. SLSN-I does not attain above-random purity or completeness under either set of priors, which may be attributed to inadequate data (SLSN-I only has six samples total).
Rare classes (Ia-91bg, Ibc, and TDE) have higher purity and lower completeness when incorporating their low frequency-based priors. 
The purity of Ia-91bg and TDE increases significantly, from $6\%\pm2\%$ to $14\%\pm5\%$ and from $8\%\pm1\%$ to $15\%\pm2\%$, respectively. Their completeness undergoes a corresponding decrease, from $13\%\pm2\%$ to $5\%\pm2\%$ for Ia-91bg and from $21\%\pm2\%$ to $9\%\pm2\%$ for TDE. The dominant Ia (unspec.) class increases in completeness from $77\%\pm1\%$ to $83\%\pm1\%$ when incorporating priors. Overall, the trends seen here reflect the expected relationship between purity and completeness, already discussed in Section \ref{sec:prob_performance} and seen in the probability-purity-completeness curves of Figure \ref{fig:pr_curves}.

\subsubsection{Anticipated Transient Identifications}
\label{sec:anticipatedrates}
In this section, we focus on our best estimate of the total number of events that may be accurately classified using our methods. These estimates, shown in Table \ref{tab:estTPFP}, correspond to the number of transient events that we anticipate being able to immediately follow-up after their initial discovery by LSST. 
\begin{table*}[]
\centering
\begin{tabular}{r|r|rrr}
\multicolumn{1}{c|}{\textbf{Class}} & \multicolumn{1}{c|}{\textbf{LSST Alerts}} & \multicolumn{1}{c}{\textbf{TP}} & \multicolumn{1}{c}{\textbf{FP}} & \multicolumn{1}{c}{\textbf{Purity}} \\
\multicolumn{1}{c|}{} & \multicolumn{1}{c|}{[$yr^{-1}$]}       & \multicolumn{1}{c}{[$yr^{-1}$]} & \multicolumn{1}{c}{[$yr^{-1}$]} & \multicolumn{1}{c}{[\%]} \\
\multicolumn{1}{c|}{} & \multicolumn{1}{c|}{}   &
\multicolumn{3}{c}{$\textrm{argmax}_{k} P(t_k |\textbf{x})$}     \\ 
\cline{2-5} 
\hline 

SNe Ia (unspec.)    & 553,277   &  458,777 & 266,333  & 63      \\
SNe Ia-91bg         & 13,398    &  678      &	4,014   & 14     \\
Ibc                 &  58,365   &  3,905    &	24,659 & 14     \\
SNe II    &  333,383  & 112,784   &  95,189   & 54     \\
SLSN-I              & 11,927    &  0        &    0      & 0       \\
TDE                 & 4,518    &  420       &	2,398   & 15     \\

\hline
\textbf{Total}        & 974,868 &  576,563 & 392,592 & -   \\
\end{tabular} 
    \caption{ Estimates of true positive (TP) and false positive (FP) identifications of LSST alerts per year for six classes. The estimates are based on the performance of our KDE multiclass classifier with frequency-based priors (Figure \ref{fig:multi_g_W2_sim_metrics}) applied to the anticipated LSST rates from PLAsTiCC \citep{Kessler_2019}. Overall, we classify about 59\% of alerts among these six classes correctly, with rates of purity ranging by class from 14\% to 63\% (excluding SLSN-I). 
    Except for SLSN-I events, we anticipate discovering at least 1-2 transients for each class per night, and thousands of the more common classes like Ia (unspec.) and II.
    }
    \label{tab:estTPFP}
\end{table*}
By accounting for the effect of redshift (\S \ref{sec:zbias}) and ensuring that our relative class frequencies are consistent with LSST (Table \ref{tab:priors}), we may assume that our KDE multiclass classifier will achieve the same purity and completeness on LSST data. We use the performance measures of the KDE multiclass classifier with frequency-based priors from Section \ref{sec:incpriors}. 

Our classification scheme requires near-infrared ($JHK$) and mid-infrared ($W1$, $W2$) magnitudes beyond the standard LSST filter set. These magnitudes are available from archival data sources, including UKIDSS, VHS, and AllWISE/unWISE. Given its unprecedented sensitivity, LSST will detect more galaxies than these shallower NIR and MIR catalogs. However, by cross-matching galaxies in the SDSS DR16 catalog (``GalaxyTag'' table) with the AllWISE and UKIDSS-LAS DR9 catalogs, we find that 50\% of SDSS galaxies brighter than r$\sim$21.34 have complete $W1/W2$ magnitudes from AllWISE and $JHK$ magnitudes from UKIDSS. We estimate that there are $\sim$5100 galaxies per square degree brighter than this magnitude limit using the magnitude distribution of SDSS galaxies in a narrow strip of the sky (R.A. = $165^{\circ}$ to $195^{\circ}$, Dec. = $+0.25^{\circ}$ to $-0.25^{\circ}$). Given that $\sim$half will have cataloged $J,H, K,W1,$ and $W2$ magnitudes, there should be $\sim$2600 galaxies per square degree with the set of magnitudes that we require for our classification. In the 18000 square degrees surveyed by LSST, there will be $\sim$46 million of these galaxies.
In the 18000 square degrees surveyed by LSST, there should be about 46 million of these galaxies. Considering the once-per-century supernova rate in typical galaxies, even our pilot methodology will provide classifications for hundreds of thousands of transients per year. Furthermore, these transients will be generally at lower redshifts where spectroscopic and multi-wavelength follow-ups are more achievable and rewarding. In this section, we assume that we will ultimately have all 19 photometric features used in our models for every LSST host.

Our classification is based only on the photometric magnitudes and colors of galaxies, without size, shape, or other morphological features. We choose this approach for our pilot program, because shape parameters are not available or uniformly measured for host galaxies in our database. Magnitudes and colors, in contrast, are generally coherently measured across different surveys and are less affected by the choice of measurement methods. Therefore, we adhere to magnitudes and color parameters in this initial work on the classification problem. Analyses of early LSST imaging will generate homogeneous samples of galaxy shape parameters (see Juri{\'c} et al. 2021\footnote{\url{https://lse-163.lsst.io/v/v3.6/index.html}}) that can easily be added as features to our models.

The PLAsTiCC test dataset reflects the anticipated number of events in each class detected by LSST over the course of three years:  1659831 Ia, 40193 Ia-91bg, 1000150 II, 175094  Ibc, 35782 SLSN-I, and 13555  TDE \citep{Kessler_2019}. We divide each of these by three to infer the anticipated LSST yearly rate (the first column of Table \ref{tab:estTPFP}). 
Modest changes in class frequencies should not significantly change completeness and purity of a probabilistic-based classifier, as verified by Figure~\ref{fig:ztest}. Hence, we use the completeness of our classifier (Figure \ref{fig:multi_g_W2_sim_metrics}) to estimate the number of events that we will be able to identify each year:
\begin{equation}
	LSST_{TP, i} = c_i  r_i ,
	\label{eq:tprate}
\end{equation}
where $c_i$ is the completeness of class $i$ from our KDE multiclass classifier, and $r_i$ is the anticipated number of samples for class $i$ among the LSST detections per year. $LSST_{TP, i}$ is thus the number of true positive detections that our model provides for class $i$ for LSST per year.  This follows from the basic idea that we are able to capture X\% of all events of class $c_i$, and given that there are $r_i$ of these events total, we will accurately capture $c_i r_i$. To determine the false positive rate, we assume THEx purity ($purity_i$) well approximates the purity we expect for LSST ($purity_i={LSST_{TP,i}}/({LSST_{TP,i} + LSST_{FP,i}})$). 
As in the case of completeness, modest differences in class frequencies entail only second order effects, which is supported by Figure~\ref{fig:ztest}. We solve for $LSST_{FP,i}$ using the purity achieved by our KDE multiclass classifier with frequency-based priors (Figure \ref{fig:multi_g_W2_sim_metrics})  and   $LSST_{TP, i}$ from Equation \ref{eq:tprate}:
\begin{equation}
	LSST_{FP, i}  =  \frac{LSST_{TP,i}}{{purity}_i} - LSST_{TP, i} .
	\label{eq:falseposrate}
\end{equation}

When considering simply the maximum assigned probability per host, we correctly classify 59\% of the alerts for these six classes. For the most dominant classes, we estimate identifying 458777 SNe Ia (unspec.) per year using an estimated completeness of about $83\%$ (top right-hand blue bar in Figure~\ref{fig:multi_g_W2_sim_metrics}) at a purity of 63\% (top left blue bar in Figure~\ref{fig:multi_g_W2_sim_metrics}). For SNe II, we estimate identifying 112784 at a purity of 54\%. Assuming the TPs are uniformly distributed throughout the year, we may expect to accurately identify 1256 Ia (unspec.), 1-2 Ia-91bg, 10 Ibc, 309 II, and 1-2 TDE each night.  These six classes represent 94\% of all extragalactic transient alerts from LSST. Uncertainties in these TP/FP estimates stem from the contamination of purity from the additional 6\% of unhandled alerts and the assumption of redshift distribution not affecting performance. The frequency of false positives may be reduced in practice by examining the entire probability distribution (examples 9-12 in Figure \ref{fig:exampleoutputs}). Viewing the probability distribution for an example galaxy may help identify false positives (e.g., galaxies that are difficult to classify may have low probabilities across all classes). 

Based on these TP/FP predictions, we can estimate a rate of correct classifications. For example, we anticipate 1-2 true positive TDE predictions per night. Given a purity of 15\%, this implies we must observe six false positives for each true positive TDE. Each night, we expect 7-8 positive predictions. Thus, we can reasonably expect one true positive TDE each night, for seven observations of positive TDE predictions. 
This is significantly better than random guessing, as TDE make up only 0.4\% of alerts among these six classes.
Thus, when our model is applied to LSST data, the resulting TDE discovery rate could exceed the current rate of tens per year.
Each night, we can expect 1-2 true positives for Ia-91bg, 10-11 Ibc, 309 II, and $>1,000$ Ia (unspec.).

\section{Conclusions}

The proposed methods presented here address the gap in probabilistically classifying transient classes using readily accessible host galaxy photometric data. By using host galaxy data, we are able to ``classify'' transient types even before they occur, enabling immediate follow-up of detected events, instead of waiting days or weeks for classification via light curves. Whereas previous approaches using host galaxy data focus on distinguishing only a handful of transient classes, or using generally unavailable host data like morphology, metallicity, or stellar mass, we minimize the host galaxy features required, enabling classification across as many galaxies as possible. 

We consider performance measures from three classifiers---binary, OVA, multiclass---for each transient class (Figures \ref{fig:binary_g_W2_metrics} and \ref{fig:multi_g_W2_metrics}) to allow the community to use the classifier that best fits their research priorities and objectives (e.g., Figure \ref{fig:exampleoutputs}).
We evaluate our methods on one of the largest collections to date of transient-host galaxy pairs \citep{Qin2021} and are able to accurately distinguish among multiple transient types using only 19 host galaxy features (10 optical-IR apparent magnitudes and nine associated colors). For example,  disregarding the relative differences in transient class frequency, we distinguish eight transient classes at balanced purities significantly above random with our KDE multiclass classifier:
Ia (unspec.), 
Ia-91bg,
Ia-91T,
Ib (unspec.),
II (unspec.),
II P,
IIn,
and TDE  (Figure \ref{fig:multi_g_W2_metrics}). We prioritize the purity of our classifications, so that
observational follow-up is efficient and produces real discoveries.

We are able to attain even higher rates of balanced purity when considering only those events that were assigned probabilities $\geq 90\%$ (Figure \ref{fig:pc_merged_all}). For the multiclass classifier, the balanced purity is significantly improved compared to  $P_M\geq 0\%$ for five classes: Ia (unspec.), Ia-91bg, II (unspec.), IIn, and TDE (Figure \ref{fig:pr_curves}).
All three classifiers often achieve a higher rate of purity at a particular probability threshold than the average purity.

Our study focuses exclusively on applying our methods to widely available photometric data so that we may predict the transient type of a significant portion of galaxies observed by LSST, using currently available galaxy data or the data collected by LSST itself. We conduct an additional analysis in Section \ref{sec:Zsection} to ensure our model is relevant to the LSST data, which is distributed differently with respect to redshift. We compare the performance of our data, as-is, to data with redshift distributions resembling those anticipated of LSST (Figure \ref{fig:zdistssampled}). We determine that there is no loss in performance for Ia (unspec.) and II (unspec.) for LSST-like test data.

Finally, we estimate the anticipated number of true positive and false positive LSST predictions that our model would provide for six transient classes: Ia (unspec.), Ia-91bg, Ibc, II (unspec.), SLSN-I, and TDE (\S \ref{sec:anticipatedrates}). These six classes consist of 94\% of all extragalactic transient alerts from LSST. We use the expected class frequencies based on the PLAsTiCC test dataset and the purity and completeness of our KDE multiclass classifier with frequency-based priors. Our method may correctly classify 59\% of the 974868 alerts per year. Each night, we anticipate accurately identifying 1-2 true positives for Ia-91bg and TDE, 10-11 Ibc, 309 II, and $>1000$ Ia (unspec.), at rates of purity ranging from 14\% to 63\%, depending on class. In practice, the number of false positive predictions may be reduced by considering the probabilities assigned to events. 
 
By using known host galaxies and corresponding transient types, we are able to train a model that can predict the class of potential transients in galaxies that have not yet hosted events. Despite the novelty of our dataset, it is still limited in its range, frequency, and completeness of classes. As new transients are observed and their host galaxies are incorporated into the dataset, the corresponding performance of the model will only improve.  

This pilot study establishes the capability for transient classification using limited host galaxy photometric data and the potential for providing immediate follow-up of transient events detected by the Rubin Observatory/LSST. Furthermore, the host galaxy-transient connections implied here are not only tools for classification, but also are themselves of astrophysical interest. Machine learning models like ours can be interrogated to find those host galaxy features most responsible for successful transient classifications, thereby illuminating the physical conditions in galaxies that may drive transient rates.

\begin{acknowledgments}
We appreciate the feedback from Decker French and Peter Behroozi. 
AIZ and Chia-Lin Ko acknowledge support from
NASA ADAP grant \#80NSSC21K0988.
AIZ also thanks the hospitality of the Columbia Astrophysics Laboratory at Columbia University, where some of this work was completed.
This work was partially funded by the University of Arizona Data Science Institute (Data7) from the Technology and Research Initiative Fund (TRIF) initiatives provided by the taxpayers of the State of Arizona. 
The software produced here uses Python 3.8.7 \citep{van1995python}, and the following packages: NumPy \citep{harris2020array}, Pandas \citep{mckinney-proc-scipy-2010}, Matplotlib \citep{Hunter:2007},  scikit-learn \citep{scikit-learn}, hmc \citep{hmc}, Jupyter \citep{Kluyver2016jupyter}, Astropy \citep{astropy:2018}, and SciPy \citep{2020SciPy}. 
\end{acknowledgments}

\appendix
\restartappendixnumbering

\section{Kernels and bandwidths}
\label{sec:appendix_kernel_bandwidth}

\begin{table}[h]
\centering  
\begin{tabular}{r|rr|rr|r}
\multicolumn{1}{c|}{\textbf{Class}} & \multicolumn{2}{c|}{\textbf{Binary}}  & \multicolumn{2}{c|}{\textbf{OVA}}  & \multicolumn{1}{c}{\textbf{Multiclass KDE}} \\ 
\multicolumn{1}{c|}{ } & \multicolumn{1}{c}{Gaussian}  & \multicolumn{1}{c|}{Exponential} & \multicolumn{1}{c}{Gaussian} & \multicolumn{1}{c|}{Exponential}  & \multicolumn{1}{c}{Exponential} \\ 
\hline
Unspecified Ia	& 0.26-0.30	    & -	            & 0.27-0.31 	& -	            & 0.041 \\
Ia-91T	        & 0.34-0.74 	& 0.14  	    & 0.36-0.57 	& 0.0001-0.12	& 0.061 \\
Ia-91bg	        & 0.28-0.32 	& 0.091-0.10    & 0.27-0.34 	& -	            & 0.041-0.082 \\
Ia-HV	        & 0.31-0.68 	& 0.14-0.33	    & 0.24-0.61 	& 0.15	    & 0.061-0.14 \\
Ic	            & 0.37-0.6 	    & 0.13-0.82	    & 0.33-0.58 	& 0.12-0.61     & 0.061-0.082 \\
Ib/c	        & 0.76       	& 0.0001-1.0    & 0.41-1.0	    & 0.0001-1	    & 0.082-0.14 \\
Unspecified Ib	& 0.3-0.79 	    & 0.081-0.6	    & 0.28-0.61 	& 0.1-0.13	    & 0.082-0.14 \\
IIb	            & 0.31-0.65 	& 0.13-0.3	    & 0.49-1.0	    & 0.1-0.49	    & 0.082-0.12 \\
Unspecified II	& 0.29-0.31 	& -	            & 0.28-0.31 	& -	            & 0.061 \\
II P	        & 0.33-0.51 	& 0.13-0.26	    & 0.29-0.45 	& 0.14-0.26	    & 0.061-0.12 \\
IIn	            & 0.37-0.65 	& 0.12-0.66	    & 0.48-0.95 	& 0.15-0.49	    & 0.061-0.12 \\
TDE	            & 0.0001-0.41 	& 0.071-0.091	& 0.31-0.4 	    & 0.1-0.15	    & 0.061-0.1 \\
\end{tabular}       
    \caption{
    Ranges of the final kernels and bandwidths per class over 10 folds from one run of 10-fold cross validation corresponding to Figures \ref{fig:binary_g_W2_metrics} and \ref{fig:multi_g_W2_metrics}.
    }
    \label{tab:kernel_bandwidth}
\end{table}

\section{Probability Performance Measures}
\label{sec:appendix_classrates}

\begin{figure}[htp]
\centering
\includegraphics[width=0.8\textwidth]{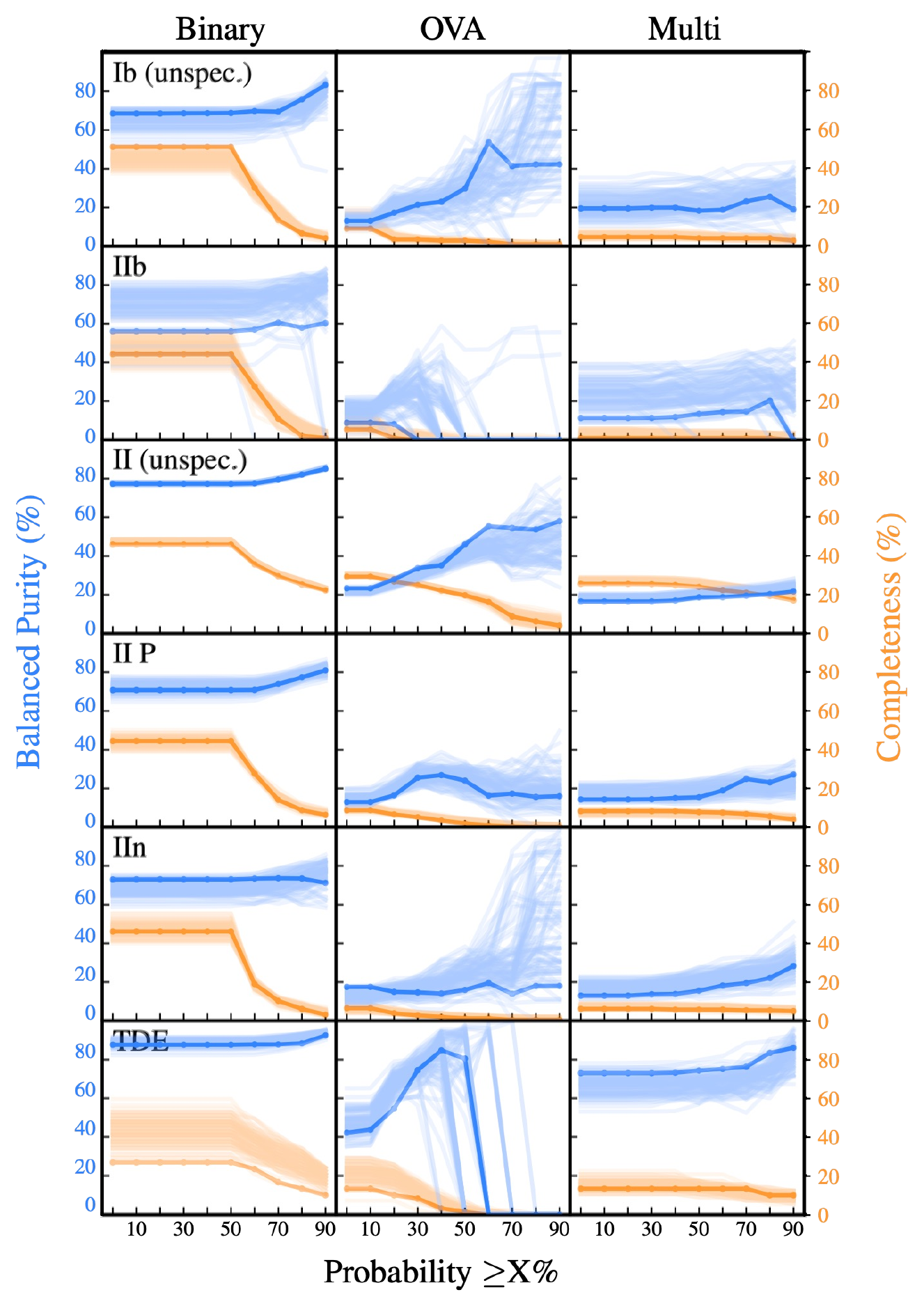}
\caption{Balanced purity (blue) and completeness (orange), as defined by Equations \ref{eq:balancedpurityrange} and \ref{eq:comprange}, versus the probability threshold assigned at $10\%$ intervals. 
This figure is the same as Figure \ref{fig:pr_curves}, but expanded to include all transient classes and all 100 trials per class. The bold lines show the first trial (which is used throughout our analysis), the fainter lines are the other 99 randomized trials.
These visualizations give an idea of the maximal balanced purity and completeness achieved per class. The binary classifiers tend to achieve the highest rates of balanced purity. 
Between the two multiclass classifiers, the KDE classifier achieves a higher maximum balanced purity than OVA for many of the rare classes: Ia-91T, Ia-91bg, Ic, IIb, and TDE. It also retains a better rate of completeness at high probabilities.
}

\label{fig:pc_merged_all}
\end{figure}

Figure \ref{fig:pc_merged_all} shows the performance improvements one can expect when utilizing the probabilities ($P \equiv P_B, P_O,$ or $P_M$) assigned to events for all three methods (binary, OVA, and multiclass) and all transient classes.
We show the balanced purity (blue lines) and completeness (orange lines) for each class for rising probability thresholds.
We consider the 11260 transient-host galaxy pairs in our dataset, dividing them into 10 subsamples, each of $\sim$10$^3$ pairs. Nine subsamples are used for training, one is used for testing. We perform 10 iterations (folds) where a different subsample is assigned as the test set each time, so that each sample is part of a test set.
To evaluate the significance of the variation in balanced purity and completeness for different probability cutoffs, we perform 100 trials over which the selection of the subsamples is randomized (faint blue and orange lines).
The bold lines are the results from the first trial run, which are also those shown in Figure \ref{fig:pr_curves} and are used throughout our analysis.

To compute the balanced purity points for Figure \ref{fig:pc_merged_all}, we compute the balanced purity using Equation \ref{eq:balancedpurity}, except that we only use data (for the true positives and false positives) at particular probability thresholds. For a particular class in a single test set, we separate samples according to the probability assigned to that target class (i.e., samples assigned a probability $\geq 10\%, \geq 20\%,$ and so on).
For each binary classifier, we count the number of true positive predictions and false positives with a probability assigned in each range. In the multiclass case, we maintain the counts for each class separately. For example, when calculating for the probability range $P_M \geq 80\%$  for Ia (unspecified), we count the total number of samples assigned a probability in the range $P_M \geq 80\%$ for Ia that are actually labeled Ia (Ia count), the total number of samples assigned a probability in the range $P_M \geq 80\%$ for Ia but that are actually of class TDE (TDE count), and so on for each class. Those measures are used in the summation of the denominator in Equation \ref{eq:balancedpurityrange}. 

For the true positives used to calculate balanced purity, we require that they were assigned the maximum probability. 
This is always the case for P $>$ 50\% as only one class can exceed the cutoff, but, for a smaller cutoff, there could be multiple candidates. For example, for a TDE cutoff of 40\%, we can have a TDE with 45\% being the maximum if all other probabilities are less than 45\%, or not the maximum if a different class has 50\%.
Altogether, the balanced purity at each threshold is calculated as:
\begin{equation}
    \textrm{BalPurityRange}(t_{k, R}) = \frac{TPR_R}{ TPR_R + \sum_{k', k' \neq k}^K \frac{FP_{k', R}}{count(k')}},
    \label{eq:balancedpurityrange}
\end{equation}
where $TPR_R=P_R / count(k)$ is the total number of samples assigned a maximum probability for class $k$ in the range ($P\geq R$) and are truly class $k$ divided by the total number of samples of class $k$ ($count(k)$). $FP_{k', R}/count(k')$ corresponds to the total number of class $k'$ samples assigned a probability in the range $R$ to class $k$ divided by the total number of samples in class $k'$ ($count(k')$). This equation allows us to compute the balanced purity at each probability threshold, for each class, as if all class sizes were equally represented. 

The completeness after applying a cutoff in an assigned probability range is
\begin{equation}
    \textrm{CompletenessRange}(t_{k, R}) = \frac{TP_R}{count(k)},
    \label{eq:comprange}
\end{equation}
where $TP_R$ is all correctly classified samples of class $k$ assigned a probability $P \geq R\%$ and $count(k)$ is the total number of samples of class $k$ in the dataset. Similar to the balanced purity, completeness is calculated relative to all samples. The true positives exclude samples when the maximum probability is outside the assigned range. Excluding those samples reduces completeness, which is shown in Figures \ref{fig:pr_curves} and \ref{fig:pc_merged_all}. The advantage of using probability cutoffs is to improve purity, which is more important than completeness for the use case of efficiently following-up LSST transients.

To find the transient classes where cutting at a high assigned probability significantly improves the balanced purity, we consider
when the averaged difference in balanced purity between $P\geq90\%$ and 0\% ($\delta_i = purity_{i, P\geq90\%}-purity_{i, P\geq0\%}$) is greater than two times the standard deviation ($\sigma$; equation \ref{eq:std}).
In Figure \ref{fig:pc_merged_all}, for the binary classifiers, we achieve a balanced purity at $P_B\geq 90\%$ significantly greater than at $P_B\geq0\%$ (by 5\%-14\%) for five transient classes: 
Ia-91bg ($8\%\pm2\%$, where $\pm$ indicates  2$\sigma$),
Ib (unspec.)($14\%\pm11\%$),
II (unspec.)($8\%\pm0.7\%$),
II P ($10\%\pm4\%$), 
TDE ($5\%\pm2\%$). 
For the rest of the classes, none exceeds two standard deviations in the negative direction except Ia (unspec.). 
For the KDE multiclass classifier, we improve the balanced purity significantly by 1\%-27\% for the following five classes: 
Ia (unspec.)($1\%\pm0.3\%$),
Ia-91bg ($27\%\pm21\%$),
II (unspec.)($6\%\pm2\%$),
IIn ($12\%\pm10\%$),
TDE ($16\%\pm15\%$). 
None produce a significantly worse result at $P_M\geq 90\%$ than at 0\%.

Sometimes the OVA classifier has no data at probabilities above 90\% so we cannot compute purity. We speculate that this apparent data-loss at high probabilities is due to the normalization of probabilities from the binary classifiers. Specifically, the lower the normalized OVA probability is, the higher the binary probabilities for competing classes must have been. For example, the binary classifier for TDE has $\sim 90\%$ balanced purity with $\sim 20\%$ completeness at $P_B\geq90\%$. However, these events may also have somewhat high probabilities assigned to other classes as well, resulting in a reduction in the probability assigned to TDE (therefore there are few events predicted as TDE with $P_O\geq60\%$). Despite this data loss at high probabilities, we still retain a similar trend of maximizing balanced purity when using probabilities assigned to events for OVA.

\begin{figure}[htp]
\centering 
\includegraphics[width=0.8\textwidth]{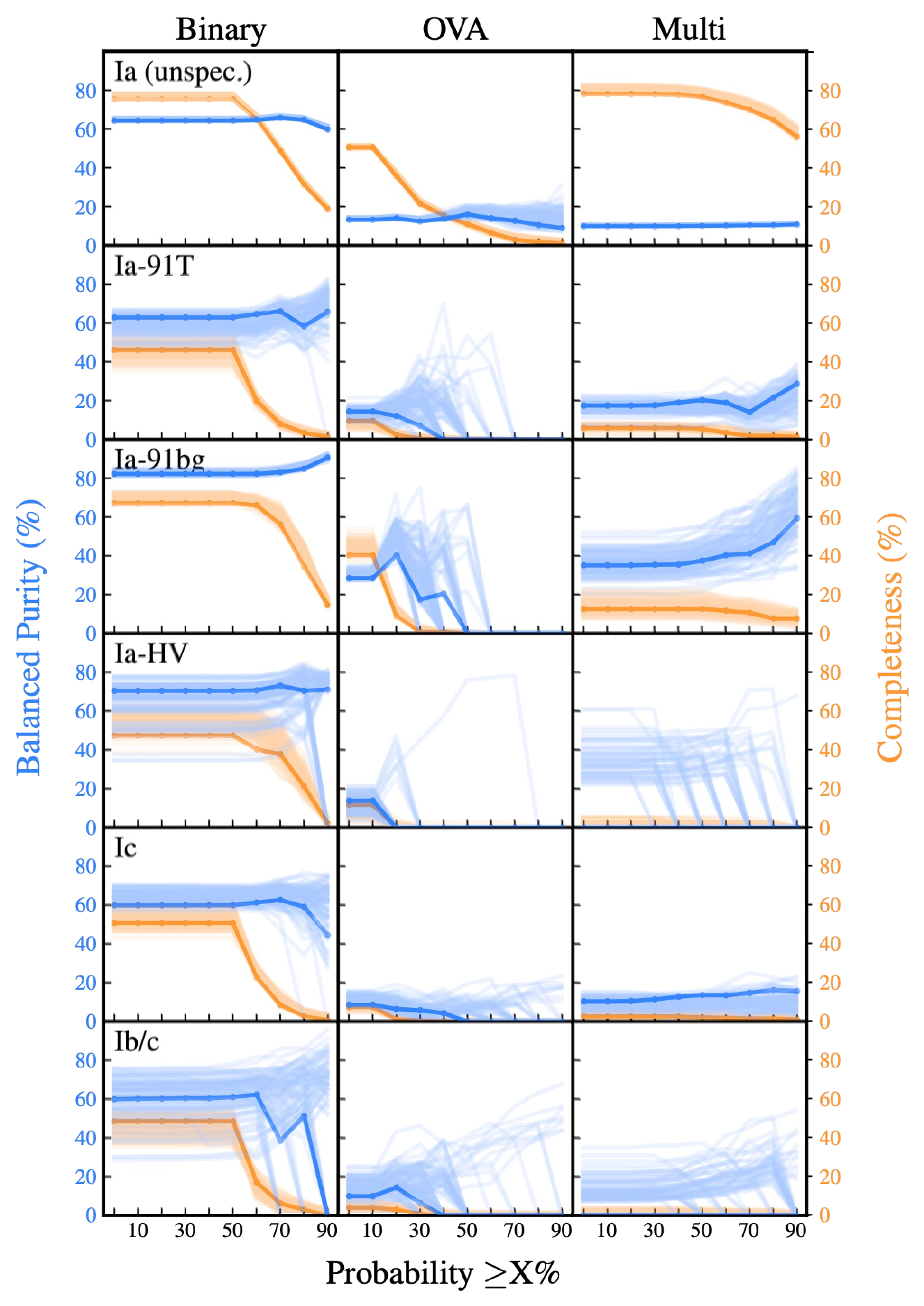}
\end{figure}
 
\bibliography{references}{}
\bibliographystyle{aasjournal}

\end{document}